\documentstyle[12pt]{article}
\renewcommand{\theequation}{\thesection.\arabic{equation}}

\newcommand{\be}{\begin{equation}}
\newcommand{\ee}{\end{equation}}
\newcommand{\bear}{\begin{eqnarray}}
\newcommand{\ear}{\end{eqnarray}}
\newcommand{\bearst}{\begin{eqnarray*}}
\newcommand{\earst}{\end{eqnarray*}}
\newcommand{\A}{\cal A}

\begin{document}
\hfill HD--THEP--01--36
\vspace{2cm}
\begin{center}\vspace*{1.0cm}
{\LARGE Cosmon dark matter?}\\
\vspace*{1.0cm}
{\large Christof Wetterich} \\
\vspace*{0.5cm}{\normalsize
{Institut f\"ur Theoretische Physik, \\
Universit\"at Heidelberg, \\
Philosophenweg 16, \\
D--69120 Heidelberg, Germany.}} \\
\vspace*{2cm}
\begin{abstract}
We investigate if the
fluctuations of the scalar field mediating quintessence -- the cosmon --
can play an important role in cosmology. Small fluctuations with short
wavelength behave similar to a relativistic gas. In contrast, the
contribution to the energy density from horizon size fluctuations may
decrease less rapidly than radiation. We discuss the possibility
that the cosmon fluctuations grow nonlinearly, form lumps and
constitute the clustering dark matter of the universe.
Cosmon dark matter would lead to interesting
consequences for the equation of state and the coupling
between quintessence and
dark matter.
\end{abstract}
\end{center}
\newpage
\section{Introduction}
The idea of quintessence was born \cite{CW1} from an attempt to
understand the vanishing of the cosmological constant. In
this work we proposed an attractor-type (``tracker'') cosmological
solution\footnote{For more details see \cite{RP}, \cite{Copeland},
\cite{CW2}.}, where a scalar field, the cosmon, ``rolls down'' an
exponential potential. This model predicts that part of the energy
density of the universe occurs in the form of homogenous ``dark
energy'' as a dynamical quantity evolving in time \cite{CW1}-\cite{ReuterW}.
In contrast to the necessity of an extreme fine tuning for a
nonzero cosmological
constant, the cosmon model implies that today's presence of the
dark energy component would be independent of details of the model
and the initial conditions. It was realized \cite{CW1} that the
late universe could become cosmon-dominated and that mild deviations
of the potential from an exact exponential form could affect the precise
present cosmological behavior. Modifications of the age of the universe
as compared to standard Friedman cosmology depend on the precise form of
the potential and the coupling of the cosmon to dark matter
\cite{CW1}, \cite{CW2}, \cite{CWV}.
Various shapes fo the potential have been investigated subsequently,
like the power law potential proposed by Ratra and Peebles \cite{RP}.

Observable cosmology, and in particular the possibility that the universe
accelerates today, depend crucially on the present effective
equation of state of quintessence \cite{ReuterW}, \cite{CL3}.
The fraction of homogenous scalar dark energy within the total
energy density, $\Omega_h$, should have been small during
structure formation \cite{CL4}. On the other hand, today's value
of $\Omega_h$ may be as large as 0.6-0.7 if one compares
the inhomogenous ``clumping'' part of dark matter $\Omega_{inhom}
\stackrel {<}{\sim} 0.4$ and the verified \cite{Boomerang} prediction
of inflation $\Omega_{tot}=1$. Combining these requirements
one concludes that $\Omega_h$ must have grown substantially
after the end of structure formation. This is realized by a
negative pressure of quintessence \cite{CL3} and leads
to a presently accelerating universe as suggested by recent supernovae
1a observations \cite{CL9}.

Present quintessence models which lead to a realistic
cosmological phenomenology\footnote{Recent bounds on the
present and past properties of quintessence can be
found in \cite{RQB}.} (at least according
to what is believed today)
seem to be somewhat in conflict with the original motivation
of quintessence, namely that today's fraction of the dark energy,
$\Omega_h(t_0)$, should not depend sensitively on the model or the
initial conditions. The power-law potentials require some tiny
parameter or ratio of mass scales which is chosen such that $\Omega_h$
plays a role just today.
Potentials with a minimum \cite{AL}, \cite{Sa} need a finetuning
of its height, location and the curvature around it.
Even the perhaps more natural modifications
of the exponential potential \cite{Hebecker} need a tuning of
parameters at the percent level which singles out indirectly
``today'' as a special moment in cosmological history.

Besides the problem of understanding the origin of quintessence
from a unified theory and its relation to the cosmological constant,
we see two major open issues in the quintessence scenario:

(1) If homogenous
quintessence accounts for 60-70 \% of the total energy density
today, what is the nature of the remaining 30-40 \% of
inhomogenous dark matter? Could it be the
inhomogenous fluctuations of the cosmon field? In this case the
cosmon would account for all dark matter (except perhaps a small
fraction in neutrinos). The space average of the cosmon field would be  
responsible for the homogenous dark energy, whereas its local fluctuations
around the average value could give rise to the inhomogenous part of dark
matter which forms structure. We may call such a form of inhomogenous
dark matter ``cosmon dark matter''. Cosmon dark matter could
give a natural explanation why it is of similar importance as
homogenous quintessence.

(2) What has triggered the recent change of the properties of quintessence
(equation of state) or its role in cosmology
such that the universe accelerates at present? Why does
acceleration happen just now?

The two questions may not be unrelated. Both may have to do with
the importance of nonlinearities for the cosmon fluctuations.
Obviously, a comparable amount of cosmon dark matter and homogenous
quintessence requires that the cosmon fluctuations are not small
at present.
Let us next ask
what  special event in the present cosmological epoch
could have modified the effective cosmological evolution
equations. A possible answer
is that structure formation has essentially taken place \cite{Hebecker}.
The existence and distribution of galaxies and clusters corresponds
to strong nonlinearities in the fluctuations of inhomogenous
dark matter. The two-point correlation function is large as
compared to the squared average energy density. Also the
higher correlation functions for the fluctuations
in the inhomogenous energy density are comparable to the
two-point function, in contrast to early cosmology. Can these
nonlinearities trigger \cite{CWBR}
the increase of $\Omega_h$ and the
acceleration of the universe by an effect of ``backreaction''?
Or can the equation of state of cosmon dark matter be negative by
itself?
An accelerating epoch would then necessarily begin only after
structure formation.

We will see that small
cosmon fluctuations on scales much smaller than the
horizon  behave essentially
as relativistic matter \cite{RP}, \cite{CL4}, \cite{Ott}.
Small-scale linear fluctuations cannot grow and can therefore not account
for structure formation. Since in the present epoch
characteristic galactic scales are much smaller
than the horizon, one concludes that the cosmon dark matter scenario can
only be realized if the nonlinearities in the cosmon
fluctuations play an important role. In this context it
should be noted that the average energy density
arising from the local cosmon field fluctuations may be distributed
much more smoothly than the individual cosmon fluctuations
during a certain period of the cosmological evolution. (This is
partly similar to the local fluctuations in the
electromagnetic field which may be substantial -- corresponding
to the photons of a relativistic gas -- and nevertheless
account for a rather homogenous energy density in radiation.)
The compatibility of an approximately homogenous cosmon energy
density (related to cosmon fluctuations and not to be confounded
with homogenous quintessence) with large average inhogeneities
in the scalar field opens two interesting alternatives. Either the cosmon
fluctuations were never linear and only the energy density
was smooth in early cosmology. Or the scalar fluctuations
started in the linear regime and have grown large in the course
of the cosmological evolution.

In this paper we concentrate on the second alternative which
is easier accessible for quantitative studies with existing
methods. Nevertheless, our results will also shed some light
on the first alternative.
One may imagine the following cosmological scenario: (1) In
an early radiation-dominated epoch of cosmology the
cosmon fluctuations and the energy
density in cosmon dark matter are small. The homogenous quintessence
would adjust to the dominant radiation
according to an attractor solution. This can lead to a
small nonvanishing $\Omega_h(t)\stackrel{<}{\sim}0.1$
during this period, compatible with nucleosynthesis. (2) For an effective
equation of state for  cosmon dark matter similar to nonrelativistic
matter, or between nonrelativistic and relativistic matter,
the part of cosmon dark matter $\Omega_c$ would grow
relative to the radiation. Cosmon dark matter could overwhelm the
radiation at a certain moment in  the history
of the universe. A cosmon-dominated epoch begins. (3)
Subsequently, structure formation could take
place by growth of the inhomogeneities of the energy density in cosmon dark  
matter. (4) Once
structure has formed, the higher correlation functions
for the energy density
of the cosmon fluctuations have also grown large. Their backreaction
could modify the effective dynamical equation for gravity and
the average cosmon
field and trigger the increase of $\Omega_h$.
In particular, the coupling of homogenous
quintessence to dark matter may
become important \cite{CW1}, \cite{Amendola}. (5) The universe
ends in a quintessence-dominated state.
We are aware that
these ideas are still speculative. We will sketch in the
following only the general setting which could describe the
cosmon dark matter scenario. Obviously, alternative solutions
to the two open issues mentioned before can be envisaged within
quintessence
\cite{Mukhanov}.

In sect. 2 we write down the field equations which couple the time
evolution of the homogenous ``average fields'' to the fluctuations.
Sect. 3 discusses the simple case of small cosmon fluctuations inside
the horizon. They behave similar to a relativistic gas. Nevertheless,
we argue in sect. 4 that the fluctuation spectrum of the cosmon needs not
to be dominated by the short wavelength fluctuations. We advocate
a natural scenario where the energy density of the cosmon fluctuations
is dominated by wavelengths of the order of the horizon -- at least
as long as they remain linear. The maximum of the spectrum becomes
therefore time-dependent.

Cosmon fluctuations with wave length around or larger than the horizon
need the inclusion of the gravity fluctuations. We discuss the
coupled linear system of fluctuation equations in sect. 5. We adopt
for this purpose the harmonic background gauge which we believe
gives a simple picture of the dominant physical processes -- in
contrast to the synchronous gauge where the growing modes
for large wavelength do not correspond to growing gauge-invariant
quantities. The growth of the cosmon fluctuations
depends on the homogenous ``background'' solution. In sect. 6 we
present several scenarios for background solutions for the specific
model of quintessence with an exponential potential. For this class
of models we compute in sect. 7 the growth exponent for the large
scale cosmon fluctuations.

In sect. 8 we turn to the radiation-dominated epoch. We establish
that the very large-scale cosmon fluctuations remain constant for a large
class of exponential cosmon potentials. However, it seems not
unnatural that the horizon size fluctuations could grow. Whereas the
contribution to the energy density from very short and very long
wavelength fluctuations decreases $\sim t^{-2}$ for the case of
constant large-scale fluctuations, the energy density
of the ``horizon component''
of the cosmon fluctuations typically could decrease somewhat slower.
As a result, the horizon component of
the cosmon fluctuations would become more important than
radiation at a certain moment in the history of the universe,
such that the radiation
epoch ends and cosmology transites to a period where cosmon dark
matter dominates. This would be a scenario where cosmon fluctuations
could grow nonlinear and form ``cosmon lumps'' -- these may
even be
possible candidates for (some) galaxies \cite{CL}.  We address
general aspects of a cosmon-dominated
universe only shortly in sect. 9 and end our discussion with a first
account of the role of the nonlinearities for the effective cosmon
fluctuation equations in sect. 10. A mean field computation shows
the tendency that non-relativistic fluctuation behavior may
shift to wavelengths further inside the horizon.
Our conclusions are presented
in sect. 11. Two appendices A, B give details for the fluctuation equation
for gravity and for the gravitational contribution to the energy
momentum tensor. For the convenience of the reader we summarize
our naming scheme for the various contributions to the
energy density in appendix C.

\section{Cosmological equations for fluctuating fields}
\setcounter{equation}{0}

We consider first only
a scalar field $\varphi$ coupled to gravity and omit radiation.
This may be a reasonable approximation for late cosmology
where photons, leptons, and baryons can be treated as a small correction.
For early
cosmology radiation will  be added later. We will work within
a formalism that is equally suited for linear and nonlinear
cosmon fluctuations. Let us start with the Einstein- and Klein-Gordon
equations\footnote{We use a metric
signature $(-,+,+,+)$ and $R_{\mu\nu\rho}^{\ \ \ \ \lambda}=
-\partial_\mu
\Gamma_{\nu\rho}^{\ \ \lambda}+...$}
in a statistical ensemble
\be\label{1}
<R_{\mu\nu}-\frac{1}{2}R g_{\mu\nu}>=\frac{1}{2M^2}
\{-<Vg_{\mu\nu}>+<\partial_\mu\chi\partial_\nu\chi>-\frac{1}{2}
<\partial^\rho\chi\partial_\rho\chi g_{\mu\nu}>\}\ee
\be\label{2}
<\chi;^\mu_{\ \mu}>=<V'>\ee
where the brackets stand for the ensemble average.
Here $V$ denotes the scalar potential and $V'=\partial V/\partial\chi$.
These field equations can be derived\footnote{Conceptually, the microscopic
action is an effective action of a quantum field theory where quantum
fluctuations are already included. A local patch of the universe obeys
the ``microscopic field equations'' which are derived by variation of eq.  
(\ref{3}) and correspond to eqs. (\ref{1}), (\ref{2}) without the brackets.
The statistical mean averages over many such patches. The ensemble therefore
consists of realizations obeying each the microscopic field equations.}
 from a ``microscopic action'' $(M^2=M_p^2/16\pi)$
\be\label{3}
S=\int d^4x\sqrt g\{-M^2R+V(\chi)+\frac{1}{2}\partial
^\mu\chi\partial_\mu\chi\}\ee
The statistical averages denoted by $<>$ depend, in principle,
on the particular observation and on the properties of the
observable local piece of the universe.
As a typical size for our ``local piece''
we may choose a redshift $z\leq 10$.
We will assume here that our local
region of the universe reflects the average properties
of a much larger region of the universe (typically extending far
beyond our horizon). In this case the statistical ensemble
should be taken as homogenous and isotropic.

For a homogenous and isotropic ensemble we expand around a
spatially flat ``background'' Robertson-Walker metric
$\overline g_{\mu\nu}(t)=<g_{\mu\nu}>=diag(-1,a^2(t),a^2(t),a^2(t))$
(with scale factor $a(t)$ and $H(t)=\dot a(t)/a(t))$ and
a background scalar field $\varphi(t)=<\chi>$, i.e.
\bear\label{4}
g_{\mu\nu}&=&\bar g_{\mu\nu}+h_{\mu\nu}\nonumber\\
\chi&=&\varphi+\delta\chi\nonumber\\
<h_{\mu\nu}>&=&0,\quad <\delta\chi>=0\ear
Insertion into eqs. (\ref{1}), (\ref{2}) yields the field equations
\be\label{5}
\overline R_{\mu\nu}-\frac{1}{2}\overline R\overline g_{\mu\nu}=
\frac{1}{2M^2}\{
-V(\varphi)\overline g_{\mu\nu}+\partial_\mu\varphi\partial
_\nu\varphi-\frac{1}{2}\partial^\rho\varphi\partial_\rho\varphi\overline  
g_{\mu\nu}+T_{\mu\nu}\}\ee
\be\label{6}
-\varphi;^\mu_{\ \mu}+V'(\varphi)=q_\varphi\ee
with energy momentum tensor
\bear\label{7}
T_{00}&=&\rho=\rho_c+\rho_g\nonumber\\
T_{ij}&=&p\bar g_{ij}=(p_c+p_g)\bar g_{ij}\ear
The energy momentum tensor and the quantity $q_\varphi$ reflect
the fluctuations\footnote{The incoherent contribution $q_\varphi$ to the
scalar field equation was pointed out in ref. \cite{CWV}.} $(i=1..3)$. Here
the scalar fluctuations lead to
\bear\label{2.8}
\rho_c(t)&=&-<V(\varphi+\delta\chi)
g_{00}>-V(\varphi)+<\dot\chi^2>\nonumber\\
&&-\frac{1}{2}<\partial^\rho\chi\partial_\rho\chi g_{00}>-\frac{1}{2}
\dot\varphi^2\nonumber\\
p_c(t)&=&-\frac{1}{3a^2(t)}<V(\varphi+\delta\chi)g_{ii}>+V
(\varphi)+\frac{1}{3a^2(t)}<\partial_i\chi\partial_i\chi>\nonumber\\
&&-\frac{1}{6a^2(t)}<\partial^\rho\chi\partial_\rho\chi g_{ii}>-
\frac{1}{2}\dot\varphi^2\nonumber\\
q_\varphi(t)&=&-<V'(\varphi+\delta\chi)>+V'(\varphi)+<\chi;^\mu_{\ \mu}>
-\varphi;^\mu_{\ \mu}\ear
whereas the pure metric fluctuations induce
\bear\label{2.9}
\rho_g(t)&=&-2M^2(<R_{00}-\frac{1}{2}Rg_{00}>-\overline R_{00}
+\frac{1}{2}\overline R\overline g_{00})\nonumber\\
p_g(t)&=&-\frac{2M^2}{3a^2(t)}(<R_{ii}-\frac{1}{2} Rg_{ii}>-
\overline R_{ii}+\frac{1}{2}\overline R\overline g_{ii})\ear
The Bianchi identity for $\overline R_{\mu\nu}$ implies the relation
\cite{CW1,CWV}
\be\label{2.10}
\dot\rho+3H(\rho+p)+q_\varphi\dot\varphi=0\ee
This equation, together with the $0-0$-components of eq. (\ref{5})
and eq. (\ref{6}),
\be\label{2.11}
H^2=\frac{1}{6M^2}(\rho+V(\varphi)+\frac{1}{2}\dot\varphi^2)\ee
\be\label{2.12}
\ddot\varphi+3H\dot\varphi+V'(\varphi)=q_\varphi\ee
specifies the evolution of the background fields. The quantities
$\rho, p$ and $q_\varphi$ express the ``backreaction'' of the
fluctuations on the effective field equations for the
mean fields\footnote{See \cite{BR} for other effects of
backreaction on quintessence.}.

The system of equations (\ref{2.10})-(\ref{2.12}) can be solved if
an ``equation of state'' $p(\rho)$ and similarly the functional
dependence $q_\varphi(\rho)$ are known. In general, the effective
functions $p(\rho)$ and $q_\varphi(\rho)$ may be quite complicated
and depend on the precise ``initial conditions''  for the
fluctuations and on time. We assume here that the system has
``locally equilibrated'' with an effective equation of state,
similar to hydrodynamcis.
Before addressing the  issue of large nonlinear
scalar fluctuations it seems useful to understand the behavior
of small fluctuations.

\section{Small cosmon fluctuations inside the horizon}
\setcounter{equation}{0}

For small fluctuations
one can expand around
the homogenous field in second order in the fluctuations. For a first
investigation we also neglect the mixing of the scalar fluctuations
with the metric fluctuations. We will see below (sect. 5) that this is
justified only if $\rho_c, p_c, q_\varphi$ are dominated by
cosmon fluctuations well inside the horizon. Nevertheless, the
approximation of inserting the background metric in eq. (\ref{2.8})
shows already several important
features in the most simple language. The fluctuations of the metric
will be included in sect. 5.
Inserting the quadratic approximations for
$V(\varphi+\delta\chi)$ and $V'(\varphi+\delta\chi)$ one finds
\bear\label{3.1}
\rho_c&=&\frac{1}{2}V''(\varphi)A+\frac{1}{2}B
+\frac{C}{2a^2}\nonumber\\
p_c&=&-\frac{1}{2}V''(\varphi)A+\frac{1}{2}B-\frac{C}{6a^2}\nonumber\\
q_\varphi&=&-\frac{1}{2}V^{(3)}(\varphi)A.\ear
In terms of the Fourier modes $\delta\chi_k$
$(\delta\chi(x)=\int\frac{d^3k}{(2\pi)^3}e^{i\vec k
\vec x}\delta\chi_k$ and $\vec x$ are
comoving coordinates) the quantities $A,B,C$ are defined as
\bear\label{3.2}
A&=&<\delta\chi^2(x)>=\int\frac{d^3k}{(2\pi)^3}\alpha_k\nonumber\\
B&=&<\delta\dot\chi^2(x)>=\int\frac{d^3k}{(2\pi)^3}\beta_k\nonumber\\
C&=&<\partial_i\chi(x)\partial_i\chi(x)>=\int\frac{d^3k}{(2\pi)^3}k^2
\alpha_k\ear
with
\be\label{3.3}
<\delta\chi^*_k\delta\chi_{k'}>=(2\pi)^3\delta(k-k')\alpha_k\ ,\
<\delta\dot\chi_k^*\delta\dot\chi_{k'}>=(2\pi)^3\delta(k-k')\beta_k\ee

Our aim is an understanding of the size and time evolution of
$A, B$ and $C$ and thereby of the equation of state for cosmon dark
matter.
For small  fluctuations the basic ingredient  is
the linearized microscopic field equation for the scalar fluctuations
(for $k\not=0$)\footnote{Note that by assumption $(q_\varphi)_{k\not=0}=0$
and $\chi_{k\not=0}=\delta\chi_{k\not=0}$. The equation for fluctuations
with $k=0$ differs from eq. (\ref{15}) if $q_\varphi\not=0$.}
\be\label{15}
\delta\ddot\chi_k+3H\delta\dot\chi_k+\frac{k^2}{a^2}\delta\chi_k
+V''\delta\chi_k=0.\ee
This can be used in order to  derive an evolution equation for $A$
\be\label{16}
\ddot A+3H\dot A+4V''(\varphi)A=-2(\rho_c-3p_c)\ee
For a given equation of state $p_c(\rho_c)$ its
solution determines $q_\varphi$. For attractor solutions with an exponential
cosmon potential one has $V''=(9/2)H^2$
and the effective cosmon mass is proportional to the
Hubble parameter \cite{CW1}, \cite{CW2}. We will, however,
keep our discussion general and consider arbitrary $V''(\varphi(t))$.

In the linear approximation there is no scattering, and the different
$k$-modes evolve independently. We may define a mode number
density $n_k$ as
\be\label{17}
n_k=\frac{1}{2}\left(\frac{k^2}{a^2}+V''\right)^{-1/2}
\left\{\left(\frac{k^2}{a^2}+V''\right)\alpha_k+\beta_k\right\}\ee
In terms of $n_k$ and the quantity $\Delta_k$
\be\label{17a}
\Delta_k=\frac{1}{2}\left(\frac{k^2}{a^2}+V''\right)^{-1/2}
\left\{\left(\frac{k^2}{a^2}+V''\right)\alpha_k-\beta_k\right\}\ee
we can write
\be\label{18}
\rho_c=\int \frac{d^3k}{(2\pi)^3}\rho_k\ ,\
p_c=\int\frac{d^3k}{(2\pi)^3}p_k,\ee
with
\bear\label{18a}
\rho_k&=&\left(\frac{k^2}{a^2}+V''\right)^{1/2}n_k,\nonumber\\
p_k&=&\frac{k^2}{3a^2}\left(\frac{k^2}{a^2}+V''\right)
^{-1/2}n_k
-\left(V''+\frac{2k^2}{3a^2}\right)\left(
\frac{k^2}{a^2}+V''\right)^{-1/2}\Delta_k\ear
For $\Delta_k=0$ eq. (\ref{2.8}) reduces to the standard relavistic
relation for a collection of point particles with mass
$\sqrt{|V''|}$ and momentum $\vec k/a$. Furthermore, for $\Delta_k
=0$ the particle number is conserved for each $k$-mode
\bear\label{19}
\frac{d}{dt}(n_ka^3)&=&(3H+\sigma)\Delta_ka^3\nonumber\\
\sigma&=&\frac{d}{dt}\ln\sqrt{\frac{k^2}{a^2}+V''}\ear
Possible unusual features are therefore all related to the
``particle production\footnote{``Mode number
production'' would perhaps be a more approriate naming.}
rate'' $\sim \Delta_k$. The latter obeys
the evolution equation
\be\label{22a}
\frac{d}{dt}(\Delta_ka^3)=(3H+\sigma)n_ka^3+2\left(\frac{k^2}{a^2}
+V''\right)^{1/2}\gamma_ka^3\ee
where $\gamma_k$ is defined by
\be\label{3.12}
\frac{1}{2}<\delta\dot \chi_k^*\delta\chi_{k'}+\delta\chi_k^*\delta
\dot\chi_{k'}>
=(2\pi)^3\delta(k-k')\gamma_k\ee
In particular, we note that in the absence of sizeable particle
production the spectrum of $n_k$ has a time-invariant shape.
Each modes scales separately according to $n_k(t)\sim a^{-3}
(t)n_k(t_0)$.

Consider first the contribution of modes with physical
wave length well within the horizon, i.e. $k^2/a^2\gg H^2$, or
a situation where $V''\gg H^2$. For $\frac{k^2}{a^2}+V''\gg H^2$ the modes
perform damped oscillations. The oscillation period will be much
shorter than $H^{-1}$ or the rate of change of $V''$.
In leading order one may therefore neglect
the damping term $\sim 3H\delta\dot\chi$, the change in $a$
and the time variation of $V''$ in eq. (\ref{15}).
One infers the frequencies $\omega^2_k=k^2/a^2+V''$. After averaging over  
initial conditions typical for the ensemble, this leads to
\be\label{3.13}
\beta_k=\left(\frac{k^2}{a^2}+V''\right)\alpha_k\ ,\quad
\Delta_k=0\ee
The damping corrections to eq. (\ref{3.13}) are suppressed by
powers
of $H^2\left(\frac{k^2}{a^2}+V''\right)^{-1}$.
This also holds for corrections from the time evolution of $V''$.
In particular, for modes
with $k^2/a^2$ much larger than $V''$ and $H^2$
the corrections decrease with time
$\sim H^2a^2$. (If the energy density has once be dominated by
such ``oscillating modes'', this will remain so later.)

For $\Delta_k=0$ the mode number density $n_k$ is directly
related to $\alpha_k$
\be\label{25a}
n_k=\left(\frac{k^2}{a^2}+V''\right)^{1/2}\alpha_k\ee
and one infers
\be\label{25b}
\rho_k=\left(\frac{k^2}{a^2}+V''\right)\alpha_k\quad,\quad  
p_k=\frac{k^2}{3a^2}\alpha_k\ee
This implies the equation of state
\be\label{22}
p/\rho=\frac{1}{3}(1+\frac{a^2V''A}{C})^{-1}\ee
\be\label{23}
q_\varphi=-\frac{V^{(3)}(\varphi)}{2V''(\varphi)}
(\rho-3p)\ee
We note that for an exponential potential $V=V_0\exp(-\alpha\varphi/M)$
and for a constant equation of state
\be\label{27A}
p=\frac{n-3}{3}\rho\ee
one can write \cite{CW2}
\be\label{27B}
q_\varphi=\frac{\beta}{M}\rho\ee
with
\be\label{3.20}
\beta=\frac{\alpha}{2}(4-n)\ee
(Cosmologies with $\beta\not=0$ have been discussed extensively
in \cite{CW2,Amendola}).

We observe that for $V''\ll C/(a^2A)$ the scalar fluctuations behave
like relativistic matter $(p\approx \rho/3)$ with $q_\varphi\approx 0$.
On the other hand, if the mass term dominates, $V''\gg C/(a^2A)$, one
finds nonrelativistic behavior $p\approx 0$ with $q_\varphi\approx-
V^{(3)}\rho/(2V'')$.
We typically concentrate on cosmon potentials where $V''$ decreases
with time faster than $a^{-2}$. (In particular, for $V''\sim t^{-2}$ and
a radiation-dominated universe $(a\sim t^{1/2}$) or matter-dominated
universe $(a\sim t^{2/3})$ one has $V''a^2\sim t^{-1}$ or
$t^{-2/3}$, respectively.) Then an individual mode with given
wavelength $\sim k^{-1}$ has the tendency to end in the
relativistic regime for large enough $t$.

In order to proceed further with
a computation of the cosmon equation of state,
we need information about the initial fluctuation
spectrum and its time evolution. As an
illustration let us first briefly consider fluctuation
spectra for which $A/C$ remains essentially constant or varies slowly
as compared to $V''a^2$.
In this case the equation of state for cosmon dark matter
approaches a relativistic gas for large $t$. Indeed, the
relativistic modes $(\frac{k^2}{a^2}\gg V'')$ remain relativistic
in the course of the
evolution, and nonrelativistic modes $(\frac{k^2}{a^2}\ll V'')$ become
relativistic at a later stage. If the cosmon ``energy spectrum''
$k^3\rho_k$ would be peaked around a fixed
comoving wave number $\overline k
\not=0$ (as for the microwave background radiation), the cosmon dark
matter would behave like relativistic matter after some time
$t_r$, where $\overline k^2/a^2(t_r)\approx V''(t_r)$. If, furthermore,
$t_r$ is in our past, such cosmon dark matter could not be relevant today.
For typical quintessence models the squared mass $V''$ is of the order
of $H^2$. Then ``linear cosmon dark matter'' (small fluctuations)
behave like relativistic matter if they are dominated by
fluctuations with wavelength well within the horizon. In this case
the cosmon dark matter may constitute a certain fraction of the
radiation \cite{Ott}. This gives a stationary contribution to $\Omega$ during
the radiation-dominated epoch. Just as for the rest of the radiation
this component will become insignificant during a ``matter-dominated
period''. In particular, it cannot play the role of dark matter and
cannot be responsible for the growth of structure.

In the following we will turn to the more interesting scenario where
the cosmon fluctuations with
wavelength around and larger than the horizon play an important role.
Indeed, since the cosmons interact only gravitationally with ordinary
matter, it seems unlikely that the cosmon fluctuations
are produced thermally. There is therefore no reason
for a peak in $k^3\rho_k$ at some fixed
$\overline k\not=0$ (as for thermally produced relics).
We rather suppose that the cosmon fluctuations have been produced
out of equilibrium at a very early stage of the cosmological
evolution, typically during inflation.

For the modes with $k^2/a^2\ll V''$,
the neglection of gravity is actually not a valid
approximation. For an appreciation of the
gravitational effects discussed in sect. 5 we will
nevertheless briefly mention here the consequences
of their omission. With $H=\eta t^{-1}$,
$V''+\frac{k^2}{a^2}=\zeta_k t^{-2}$ and $\tau=\ln(t/t_0)$
the long-distance modes obey the field
equation
\be\label{3.21}
\frac{d^2}{d\tau^2}\delta\chi_k+(3\eta-1)\frac{d}{d\tau}\delta  
\chi_k+\zeta_k\delta\chi_k=0\ee
If the $\tau$-dependence of $\eta$ and $\zeta_k$ can be neglected
(i.e. $V''$ dominates $\zeta_k$), the
solution reads (with $\omega_k,\gamma_k$ real)
\be\label{L1}
\delta\chi_k(\tau)=e^{i\omega_k\tau}\ e^{-\gamma_k\tau}\delta\chi_k(0)\ee
with
\be\label{L2}
\gamma=\frac{3\eta-1}{2}\quad,\quad \omega^2_k=\zeta_k-\frac{(3\eta-1)^2}
{4}\ee
This leads  to
\be\label{L3}
\beta_k=t^{-2}(\omega^2_k+\gamma^2)\alpha_k=\left(\frac{k^2}{a^2}+
V''\right)\alpha_k\ee
We conclude that without gravity effects the
``particle production'' $\Delta_k$ is negligible both for modes with
$k^2/a^2+V''\gg H^2$ and $\frac{k^2}{a^2}\ll V''$.
We will see below, however, that $\Delta_k$ is actually of the same size
as $n_k$ once gravity effects are included for the long-distance modes.

\section{Cosmon spectrum}
\setcounter{equation}{0}
The equation of state of cosmon dark matter depends
strongly on the spectrum of the cosmon fluctuations. In contrast
to particles that have once been in thermal equilibrium the cosmon
fluctuations are typically  produced by out
of equilibrium physics in very early cosmology (e.g. inflation).
We have  little a priori knowledge about the primordial spectrum
without understanding the role of the cosmon in this very early
epoch. Nevertheless, the spectrum is also reshaped (``processed'')
by the different time evolution for modes with different $k$.
As a result, we will see that a spectrum with a maximum
in the vicinity of the momentum corresponding to the horizon appears
to be rather natural.

Let us parametrize our lack of knowledge by a general form
of the spectrum
\be\label{S.1}
\alpha_k(t)=\bar\alpha(t)M^2k^{\nu_{IR}-3}(1+k/k_p(t))
^{\nu_{UV}-\nu_{IR}}\ee
For small and large $k$, it has the properties
\be\label{AD0}
k^3\alpha_k\sim\left\{\begin{array}{lll}
k^{\nu_{IR}}& {\rm for}& k\ll k_p\\
k^{\nu_{UV}}&{\rm for}& k\gg k_p\end{array}\right.\ee
This spectrum leads to a finite energy density if
the power laws for small and large $k$ obey

\be\label{S.2}
\nu_{IR}>0\quad,\qquad\nu_{UV}<-2\ee
It is peaked around a characteristic comoving momentum
$k_p(t)$ which may depend on time. Performing the integrals
(\ref{3.2}) one finds
\be\label{S.3}
A={\cal A}(t)M^2k_p(t)^{\nu_{IR}}\quad, \quad C=c_pk^2_p(t)A\ee
The generic form of eq. (\ref{S.3}) holds for a wide class
of spectra with scaling properties similar to eq. (\ref{S.1}),
whereas the constant $c_p$ depends on the particular shape
and reads in our case
\be\label{S.4}
c_p(\nu_{IR},\nu_{UV})=\frac{\Gamma(\nu_{IR}+2)\Gamma(-\nu_{UV}-2)}
{\Gamma(\nu_{IR})\Gamma(-\nu_{UV})}\ee
We may distinguish three types of qualitatively
different behavior:
\begin{enumerate}
\item[i)]
horizon-dominated spectrum: in this case the physical momentum
corresponding to the peak of the spectrum scales with the horizon, i.e.
\be\label{S.5} k_p(t)=\kappa H(t)a(t)\ee
\item[ii)] radiation-like spectrum: here $k_p(t)$ is a fixed
comoving wave number with corresponding physical length scale well
within the horizon, such that $k_p/a(t)\gg H(t)$.
\item[iii)]
infrared spectrum: it is dominated by wave lengths outside the
horizon, with $k_p/a(t)\ll H(t)$.
\end{enumerate}

The normalization ${\cal A}(t)$ is proportional to $\bar\alpha(t)$.
It is treated here as a free parameter. We remind, however,
that the cosmon fluctuations have a direct impact on the fluctuations
in the cosmic microwave background and are therefore restricted
in amplitude at the time of last scattering.
In fact, the microscopic fluctuations of the cosmon dark energy
$\delta\rho_k$ gets a contribution that is
linear in the fluctuations $\delta\chi_k,\delta\dot
\chi_k$. This is related to the fact that the cosmon does not oscillate
around a constant value, in contrast to photons and  baryons.
For example, the contribution to the energy density $\sim V'\delta\chi$
does not vanish since the background field is not at a minimum of
$V$ and therefore $V'\not=0$.
As a consequence, the metric fluctuations
$(h_{\mu\nu})_k$ and therefore also the
temperature fluctuations $(\delta T/T)_k$ in the cosmic
microwave spectrum get a contribution $\sim\alpha_k,\beta_k$. It is
conceivable, nevertheless, that the proportionality constant
is rather small such that rather large cosmon fluctuations induce only
small fluctuations in the radiation.

For the spectrum (\ref{S.1}) we can determine the evolution
of ${\cal A}(t)$ from the evolution of the modes with momenta
smaller than $k_p$
\be\label{S.6}
\partial\ \ln {\cal A} (t)/\partial\ \ln t=\partial\ln\alpha_{k\ll k_p}(t)
/\partial\ \ln t=f_<\ee
On the other hand, the combination ${\A} k_p^{\nu_{IR}-\nu_{UV}}$
scales with the modes $k\gg k_p$
\be\label{S.7}
\partial\ \ln\ {\A}(t)/\partial\ \ln t+(\nu_{IR}-\nu_{UV})\partial\ \ln
k_p(t)/\partial\ \ln t=\partial\ \ln \alpha_{k\gg k_p}(t)
/\partial\ \ln t=f_>\ee
Knowledge of the time evolution for the modes
on both sides of the peak of the spectrum is therefore
sufficient to determine both ${\A}(t)$ and $k_p(t)$. (This holds
as long as the parametrization (\ref{S.1}) remains valid.)
Indeed, one finds the simple relation
\be\label{AD1}
\frac{\partial\ \ln k_p}{\partial\ \ln t}=\frac{f_>-f_<}{\nu_{IR}
-\nu_{UV}}\ee
Inversely, if $k_p(t)$ is known, this yields a relation for
the difference in the slopes, $\nu_{IR}-\nu_{UV}$.
For the
radiation spectrum and the infrared spectrum the growth exponent
$\partial\ln \alpha_k/\partial\ln t$ is typically the
same on both sides of the peak.
Therefore $k_p$ remains constant as expected. We have already
argued in the last section that the radiation-like spectrum
indeed leads to the equation of state of relativistic matter
with $q_\varphi=0$. The radiation-like
spectrum seems not to be well motivated or interesting, however.

For the horizon-dominated spectrum
the difference in the slope of the the spectrum on both sides of the
peak is due to the different growth rates for short wavelength and
infrared modes. Typically, the short-distance modes are relativistic,
whereas the modes with $k$ somewhat smaller than $k_p$ are nonrelativistic.
The growth rate of the infrared modes is also influenced
by the gravitational interaction and depends, in general, on the
chosen gauge of the metric. If $k_p$ scales with the horizon, $k_p\sim
aH$, one has a general connection between the growth rates $f_>,f_<$ and
the difference in slope between the ultraviolet and infrared
spectrum $(H=\eta t^{-1})$
\be\label{AD2}
\frac{\partial\ \ln k_p}{\partial \ln t}=-(1-\eta),\ee
\be\label{AD3}
\nu_{IR}-\nu_{UV}=\frac{f_<-f_>}{1-\eta}\ee
A spectrum with finite energy density $(\nu_{IR}-\nu_{UV}>2)$
requires $f_<-f_>>2(1-\eta)$.

For the purpose of later comparison,
we first neglect gravity and incorporate it in the next sections.
Within the approximations of the last section the conserved particle
number (\ref{19}) and the relation (\ref{25a}) implies then
\be\label{S.8}
\frac{d}{dt}\left[a^3\left(\frac{k^2}{a^2}+V''
\right)^{1/2}\alpha_k\right]=0\ee
such that
\be\label{S.9}
\partial_t\ln\alpha_k=-3H-\frac{1}{2}\partial_t\ln\left(\frac{k^2}
{a^2}+V''\right)\ee
One infers for $k^2_p/a^2\sim V''\sim t^{-2}$ and
$H=\eta t^{-1}$ the different growth rates
\bear
f_>&=&\partial\ln\alpha_{k\gg k_p}/\partial\ln t=-2\eta,\label{S.10}\\
f_<&=&\partial\ln\alpha_{k\ll k_p}/\partial\ln t=-(3\eta-1)\label{S.11}\ear
and therefore
\be\label{S.12}
\frac{\partial\ln k_p}{\partial \ln t}=-\frac{1}{\nu_{IR}-
\nu_{UV}}(1-\eta)\ee
On the other hand, the separation between relativistic and nonrelativistic
behavior at $k^2_p\sim V''a^2$ corresponds to (\ref{AD2})
and we conclude that the faster decay of the relativistic modes as
compared to the nonrelativistic ones leads to a spectrum with
\be\label{S.14}
\nu_{IR}-\nu_{UV}=1\ee
This difference is not enough to fulfil simultaneously both bounds
in eq. (\ref{S.2}).

Actually, the relation (\ref{S.8}) is only obeyed for the modes
well within the horizon and only
the estimate of $f_>=-2\eta$ is therefore reliable. We may
combine the relations (\ref{AD2}),
(\ref{S.7}), (\ref{S.6}) and the growth rate for the relativistic
modes (\ref{S.10}) to infer
\be\label{S.15}
f_<=\partial\ln \alpha_{k\ll k_p}/\partial\ln t=(1-\eta)(\nu_{IR}
-\nu_{UV})-2\eta\ee
Consistency with the bounds (\ref{S.2}) requires then $\nu_{IR}-\nu_{UV}
>2$ or
\be\label{S.16}
f_<>2-4\eta\ee
We observe that for $\eta=\frac{1}{2}$
(as in the radiation-dominated epoch) a consistent picture
for the spectrum (\ref{S.1})
emerges only if the long wavelength modes grow, i.e. $f_<>0$.
We will
see in the next section how gravity effects induce a nonvanishing
particle production rate $\Delta_k$ and  modify the growth
rate $f_<$ as compared to eq. (\ref{S.11}). Depending on the shape
of the cosmon potential, we will see in sect. 8 that
indeed $f_<$ is zero during the radiation-dominated epoch.

Finally, we discuss for the spectrum (\ref{S.1}) the time evolution of the  
quantities $A$ and
$C$ for arbitrary $f_<$. One has for a horizon-dominated spectrum
(cf. eq. (\ref{S.3}), (\ref{AD2}))
\be\label{S.17}
\frac{\partial\ln A}{\partial \ln t}=f_<-(1-\eta)\nu_{IR}
=-(1-\eta)\nu_{UV}-2\eta\ee
and $A$ can only grow for positive $f_<>(1-\eta)\nu_{IR}$.
The relevant quantity for the equation of state
\be\label{S.18}
\frac{C}{a^2H^2A}=c_p\kappa^2\ee
is time-independent (for $k^2_p/a^2=\kappa^2H^2$).
The relation (\ref{S.18})
also  shows that for a horizon-dominated spectrum the equation
of state for cosmon dark matter will depend on the details
of the spectrum around the maximum since the constants $c_p$ and $\kappa$
enter crucially.

Quite generally the details around the maximum
of $k^3\alpha_k$ depend crucially on the precise
understanding of the evolution of the modes with
momenta near the maximum. On the other hand, the relation
(\ref{AD1}) between the slopes in the infrared and ultraviolet
is insensitive to these details. This may be demonstrated
by assuming a different growth rate and different slopes
of the spectrum in three different regions, $(k_1\ll k_2)$
\be\label{AD4}
k^3\alpha_k\sim\left\{
\begin{array}{lll}
k^{\nu_{IR}}& {\rm for}& k\ll k_1(t)\\
k^{\nu_{max}}&{\rm for}& k_1(t)\ll k\ll k_2(t)\\
k^{\nu_{UV}}& {\rm for}& k\gg k_2(t)\end{array}\right.\ee
as exemplified by
\be\label{AD5}
\alpha_k=\bar\alpha(t)M^2k^{\nu_{IR}-3}\left(1+\frac{k}{k_1(t)}\right)^
{\nu_{max}-\nu_{IR}}
\left(1+\frac{k}{k_2(t)}\right)^{\nu_{UV}-\nu_{max}}
\ee
Defining the growth rates $f_<$ and $f_>$ for
the infrared and ultraviolet modes as before (e.g. eqs. (\ref{S.10}),
(\ref{S.11})
with $k_p$ replaced by $k_1, k_2$)
and
\be\label{AD6}
f_{max}=\partial\ \ln \alpha_{k_1\ll k\ll k_2}/\partial\ \ln t\ee
leads to the relations
\bear\label{AD7}
f_{max}-f_<&=&(\nu_{IR}-\nu_{max})\frac{\partial\ \ln k_1}{\partial\ \ln t},
\nonumber\\
f_>-f_{max}&=&(\nu_{max}-\nu_{UV})\frac{\partial\ \ln k_2}{\partial\ \ln t}
\ear
If both $k_1$ and $k_2$ scale proportional to $k_p$, we indeed
recover the relation (\ref{AD1}) independent of the growth
rate of the modes near the maximum, $f_{max}$.
This tells us that in the case where $f_<$ does not obey the bound
(\ref{S.16}) another scale different from the horizon
must be present in the spectrum. One possibility could be
an effective ultraviolet cutoff $k_{UV}$ beyond which the
spectrum is further suppressed. Due to the extremely weak
interaction of the cosmon with ordinary matter such a cutoff
does not arise from microscopical processes in the late universe.
A natural scale for such a cutoff could be the scale which corresponds
to the horizon at the end of inflation. In this case $k_{UV}$ does
not depend on time. (Today $a/k_{UV}$ is then typically in the
centimeter range.)

\section{Linear gravity effects}
\setcounter{equation}{0}

We have seen in the preceeding section that a crucial ingredient
for the understanding of the spectrum of the cosmon fluctuations
and therefore also for the equation of state of cosmon dark matter
is the understanding of the time evolution of the modes
with wavelength around and outside the
horizon. For these modes the gravitational effects become important.
Indeed, the effects of the gravitational attraction become comparable
in strength to the effects of the mass term $V''$ which
scales typically with the Hubble parameter, $V''\sim H^2$. We
therefore
include next the effects of gravity in the linear approximation.
We adopt here the ``background harmonic gauge''
\be\label{S.AA}
D_\mu h^\mu_{\ \nu}=\frac{1}{2}\partial_\nu h\ee
where $D_\mu h^\mu_{\ \nu}=\partial_\mu h^\mu_{\ \nu}+(
\bar g^{-1/2}\partial_\mu\bar g^{1/2})h^\mu_{\ \nu}
-\bar\Gamma_{\mu\nu}^{\ \ \lambda}h^\mu_{\ \lambda}$ is
the covariant derivative in the background metric $\bar g_{\mu\nu}$
which is also used to raise and lower the indices of $h_{\mu\nu}$.
It has the property that in linear order $(\delta g_{\mu\nu}=h_{\mu\nu})$
\be\label{S.AA2}
\bar g^{\mu\nu}\delta\Gamma_{\mu\nu}^{\ \ \lambda}=0\ee
and therefore
\be\label{S.AA3}
\delta(\chi;^\mu_{\ \mu})=(\delta\chi);^\mu_{\ \mu}-h^{\mu\nu}\varphi;
_{\mu\nu}\ee
We believe that the background harmonic gauge gives us
a comparatively simple picture of the physical situation. This is
crucial for an understanding of the large-scale cosmon fluctuations
and justifies the somewhat higher algebraic complexity as compared
to the more usual synchronous gauge. (For the synchronous gauge
the growth of the dominant mode for large $t$ does not
correspond to a growing gauge-invariant fluctuation and obscures
the picture.)

The gravity effects result
in a modification of the microscopic evolution equation
for the cosmon fluctuations\footnote{Note that one has
to linearize precisely the equation $(\chi;^\mu_{\ \mu}-V')_k
=\delta(\chi;^\mu_{\ \mu})_k-(\delta V')_k-
(q_\varphi)_k=0$, using $(q_\varphi)_{k\not=0}=0$.)} for $k\not=0$
\be\label{LG1}
\delta\ddot\chi_k+3H\delta\dot\chi_k+\left(\frac{k^2}{a^2}
+V''\right)\delta\chi_k=H\dot\varphi ((h_{00})_k+h_k)
-\ddot\varphi (h_{00})_k\ee
The field equations for the metric fluctuations are derived
in appendix A.
In harmonic gauge  the trace of the
metric fluctuations, $h=h^\mu_\mu=\bar g^{\mu\nu}h_{\nu\mu}$, obeys (for  
$k\not=0, h_{k=0}\equiv0$)
\be\label{A111}
h;^\mu_{\ \mu}=-
\frac{1}{M^2}(4V'\delta\chi-2\dot\varphi\delta\dot\chi
+Vh+\frac{\rho+p}{2}(h+4h_{00}))\ee
We observe that contributions involving the
potential of the background field $(\sim V,V')$ conserve
the Lorentz symmetry and therefore multiply Lorentz-scalars
$\delta\chi, h$. On the other hand, terms involving $\dot\varphi$,   
$\ddot\varphi$ or $\rho+p$ respect only the rotation symmetry. This explains
the separate appearance of $h_{00}$.

A second effect of gravity are corrections to the energy momentum
tensor and to $q_\varphi$ according to eq. (\ref{2.8}). Using
the definitons of (\ref{18}) and similarly $q_\varphi=\int_kq_k$, the
additional contributions $(\rho_k=(k^2/a^2+V'')^{1/2}n_k+\Delta\rho_k$
etc.) up to second order in $h_{\mu\nu}$ are
\be\label{XXC}
\Delta\rho_k=-V'<\delta\chi h_{00}>_k\ee
\bear\label{XXD}
\Delta p_k&=&-\frac{V'}{3}<\delta\chi h_i^i>_k+\frac{\dot\varphi}
{3}<\delta\dot\chi h_i^i>_k\nonumber\\
&&+\dot\varphi<\delta\dot
\chi h_{00}>_k+\frac{\dot\varphi^2}{2}<h_{00}^2>_k+
\frac{\dot\varphi^2}{6}
<h_{00}h_i^i>_k\ear
\bear\label{XXE}
\Delta q_k&=&<\chi;^\mu_{\ \mu}>_k=<\frac{1}{2}\partial^\nu h\partial_\nu  
\delta\chi+3H h_0^\nu\partial_\nu\delta\chi-\partial_\mu h^{\mu\nu}
\partial_\nu\delta\chi\nonumber\\
&&-h^{\mu\nu}\partial_\mu\partial_\nu\delta\chi>+(\ddot\varphi+3H
\dot\varphi)<h_0^{\ \rho}h_{\rho 0}>\nonumber\\
&&-\dot\varphi<\partial_\mu h^{\mu\rho}h_{\rho 0}+h^{\mu\rho}\partial_\mu
h_{\rho 0}-\frac{1}{2}\partial_\mu hh_0^\mu-\frac{1}{2}h^\rho_\nu
\dot h_\rho^\nu>\ear
We observe that the relations (\ref{XXC})-(\ref{XXE})
do not yet use a particular
gauge. Finally, the metric fluctuations lead to a pure
gravitational contribution to the energy momentum tensor $\rho_g,p_g$
(cf eq. ({\ref{2.9})). This is discussed in appendix B.

We need the relative size of $h^i_i=h_{ii}/a^2$ and $h_{00}$
with $h=-h_{00}+h_i^i$. This issue is discussed in appendix A.
For many purposes the ratio
\be\label{4.12AA}
y_k=\frac{(h_{00})_k}{h_k}\ee
can  be taken as a time-independent (complex) constant.
(This holds for $V'/\dot\varphi=const$ and
$t\delta\dot\chi_k\sim\delta\chi_k$,
with complex proportionality factor for oscillating $\delta\chi_k$.)
One obtains the microscopic equation for the
trace of the metric fluctuations
\be\label{A112}
\ddot h_k+3H\dot h_k+\left(\frac{k^2}{a^2}
-\frac{V}{M^2}-\frac{\rho+p}{2M^2}
(1+4y_k)\right)h_k=\frac{1}{M^2}(4V'\delta \chi_k-2\dot
\varphi\delta\dot\chi_k)\ee
This is our central equation for the time evolution of $h$. We note
that it continues to hold in presence of additional
radiation fluctuations which obey $\delta\rho_r-3\delta p_r=0$ and do not
contribute to $\delta t^\rho_\rho=\bar g^{\rho\lambda}
\delta t_{\rho\lambda}$. Besides the contribution
to $\rho+p$, additional radiation can therefore
only affect the value of $y_k$.

For large enough $k$ we can neglect the retardation effects
(i.e. neglect the
time derivatives of $h_k$) and approximate
\be\label{LG2}
h_k=-\frac{a^2}{M^2k^2}(\delta t^\mu_\mu)_k=\frac{4a^2V'}{M^2k^2}
\delta\chi_k-\frac{2a^2\dot\varphi}{M^2k^2}\delta\dot\chi_k\ee
The r.h.s. of eq. (\ref{LG1}) therefore becomes
\be\label{LG3}
H\dot\varphi(h_k+(h_{00})_k)-\ddot\varphi(h_{00})_k=
[H\dot\varphi+(H\dot\varphi-\ddot\varphi)y_k]\left(\frac{4a^2V'}
{M^2k^2}\delta \chi_k-\frac{2a^2\dot\varphi}{M^2k^2}\delta\dot\chi_k
\right)\ee
and the modified microscopic equation reads
\be\label{LG4}
\delta\ddot\chi_k+\left\{
3H+\frac{2u_kH\dot\varphi^2a^2}{M^2k^2}
\right\}\delta\dot\chi_k\nonumber\\
+\left\{\frac{k^2}{a^2}+V''
-\frac{4u_kH\dot\varphi V'a^2}{M^2k^2}\right\}\delta\chi_k=0\ee
with
\be\label{5.17a}
u_k=1+y_k-\frac{y_k\ddot\varphi}{H\dot\varphi}\ee
We observe a time-dependent critical value of $k$,
\be\label{LG5}
k^2_c(t)=\frac{a^2\dot\varphi^2}{M^2}=O(a^2H^2)\ee
such that for $k^2\gg k^2_c$ the gravitational corrections become
negligible\footnote{The details
of this estimate depend on the time evolution of $V',V''$.}.
For the modes in this momentum range the results
of the last section apply and the validity of the
approximation (\ref{LG2}) can be checked by
an iterative solution.
For a large class of cosmon potentials one has $\dot
\varphi^2\sim t^{-2}$ and $k^2_c$ decreases with time such that
the range of ``relativistic modes'' $k>k_c(t)$ increases.
On the other
hand, for the long wave-length modes
with $k^2\stackrel{\scriptstyle<}{\sim}
k^2_c$ the attraction by gravity becomes important. We will investigate
these modes in sect. 7.

\section{Quintessence with exponential potential}
\setcounter{equation}{0}

The details of the gravitational effects depend on the particular
quintessence model, i.e. the time history of $V',V''$ and $\dot\varphi$.
For an illustration, we take here an exponential potential
\be\label{LG6}
V(\varphi)=M^4\exp(-\alpha\frac{\varphi}{M})\ee
In addition to cosmon dark matter we also allow for an
independent radiation component from other particles. The combined
energy density and pressure
from radiation and cosmon dark matter will be
denoted by $\bar\rho=\rho_r+\rho$ and similarly
$\bar p=\frac{1}{3}\rho_r+p$. We also assume (cf. eq. (\ref{27B})
\be\label{PPX}
\bar q_\varphi=\frac{\bar\beta}{M}\bar\rho\ee
with constant $\bar\beta$.
An asymptotic solution for $3\bar p=(\bar n-3)\bar \rho$ with constant
$\bar n$ was found in \cite{CW1,CW2}
\bear\label{LG7}
H&=&\eta t^{-1},\quad \dot\varphi=\frac{2M}{\alpha}
t^{-1},\quad \ddot\varphi=-\dot\varphi t^{-1}\quad
,\quad V''=\zeta_0t^{-2}\nonumber\\
V'&=&-\frac{\alpha}{M}V=-\frac{M}{\alpha}V''=-\frac{M\zeta_0}{\alpha}
t^{-2}\ear
Such a solution requires
\be\label{P1}
\eta=\frac{2}{\bar n}(1-\frac{\bar\beta}{\alpha})\ee
and
\be\label{P2}
\zeta_0=\frac{2}{\bar n^2}\{\bar n(6-
\bar n)-12\bar\beta(\alpha-\bar\beta)\}\ee
Inserting these values yields algebraic relations for
the energy density in particles (other than the cosmon)
$\Omega_p=\bar\rho/\rho_{cr}$, the scalar
potential $\Omega_V=V/\rho_{cr}$ and
homogenous quintessence
$\Omega_h=(V+\frac{1}{2}\dot\varphi^2)/\rho_{cr}$ (with $\rho_{cr}=
6M^2H^2$)
\bear\label{P3}
\Omega_p&=&1-\frac{\bar n-2\bar\beta(\alpha-\bar\beta)}{2(\alpha-
\bar\beta)^2}\nonumber\\
\Omega_V&=&\frac{\bar n(6-\bar n)-12\bar\beta(\alpha-\bar\beta)}{
12(\alpha-\bar\beta)^2}=
\frac{\zeta_0}{6\alpha^2\eta^2}\nonumber\\
\Omega_h&=&1-\Omega_p\ear
We note that homogenous quintessence $(\rho_h,p_h)$ receives
contributions both from the potential and the kinetic
energy of the scalar background field $\varphi$.
The positivity of $\Omega_p$ and $\Omega_V$ imposes
the bounds
\bear\label{P4}
&&\alpha(\alpha-\bar\beta)>\frac{\bar n}{2}\nonumber\\
&&\bar\beta(\alpha-\bar\beta)<\frac{\bar n(6-\bar n)}{12}\ear
We will concentrate on three special cases:

\bigskip\noindent
\underline{a) Radiation domination}

Let us first consider a situation where cosmon dark matter is
subdominant as compared to the dominant radiation from
other particles. In this case one has $\bar n=4$ and $\bar\beta=0$
such that the solution exists for $\alpha^2>2$. One finds
\be\label{P5}
\eta=\frac{1}{2}\ ,\quad \zeta_0=1\ , \quad\Omega_h=
\frac{2}{\alpha^2}\ee

\bigskip\noindent
\underline{b) Cosmon dark matter dominated by large-scale fluctuations}

Next we  investigate a scenario where the inhomogenous
energy density is dominated by cosmon fluctuations
such that  $\bar n=n,\ \bar\beta=\beta$. We assume that the properties of
cosmon dark matter can be approximately described by the large-scale
cosmon fluctuations.
In this case the term $\sim C$ in eq. (\ref{3.1}) can be neglected and one
finds
\be\label{P6}
(q_\varphi-\Delta q_\varphi)=\frac{\alpha}{2M}
(\rho-p-\Delta\rho+\Delta p)\ee
Without the gravity corrections (\ref{XXC}),
i.e. for $\Delta q_\varphi=\Delta\rho=
\Delta p=0$,  this would yield
\be\label{P7}
\beta=\frac{6-n}{6}\alpha\ee
and therefore
\be\label{P8}
\eta=\frac{1}{3}\quad, \quad\zeta_0=\frac{2(6-n)}{n}
\left(1-\frac{\alpha^2}{3}\right)\ee
In this approximation the bounds (\ref{P4}) would require
simultaneously $\alpha^2>3$ and $\alpha^2<3$ and we could
consider the solution at best as a limiting case where the cosmon
kinetic energy dominates. We will argue, however, that the gravity
corrections (\ref{XXC}) need to be included and therefore the
relation (\ref{P7}) does not hold. For the time being we will consider
$\beta$ and $n$ as free parameters characterizing the incoherent
fluctuations. We will discuss in sect. 9 how  to determine them self-consistently.

\newpage\noindent
\underline{c) Cosmon domination in absence of particle production}

Finally we investigate a situation where the inhomogenous energy
density is dominated by cosmon dark matter in the absence of
particle production. This last scenario is somewhat hypothetical and may
apply in a situation where the cosmon fluctuations are already
nonlinear and wave lengths inside the horizon dominate, without
showing the characteristics of a relativistic gas. An example
may be cosmon lumps \cite{CL}.For this case one has again $\bar n=n$ and
$\bar\beta=\beta$. Omitting the corrections (\ref{XXC}) one has in absence
of particle production $(\Delta_k=0$, cf. eq. (\ref{3.20}))
\be\label{LG8}
\beta=\frac{\alpha}{2}(4-n)\ ,\quad \eta=\frac{2}{n}\left
(1-\frac{\beta}{\alpha}\right)=1-\frac{2}{n}\ee
and
\bear\label{LG8AA}
V''&=&\frac{\alpha^2}{M^2}V=\frac{2n(6-n)-6\alpha^2(4-n)(n-2)}{(n-2)^2}H^2
\nonumber\\
&=&\frac{2}{n^2}\{ n(6-n)-3\alpha^2(4-n)(n-2)\}t^{-2}=
\zeta_0t^{-2},\ear

We emphasize the unusual expansion rate for $n\not=4$ due to the coupling
$\beta$. In particular, for the equation of state of nonrelativistic
matter, $n=3$, we find $\eta=1/3$ and therefore
$a\sim t^{1/3}$ instead of the usual $a\sim t^{2/3}$. The relative  
contribution to the energy density of cosmon dark matter is given for this  
solution
by
\be\label{LG8a}
\Omega_p=\frac{2}{n-2}\left(1-\frac{n}{\alpha^2(n-2)}\right)\ee
whereas homogenous
quintessence accounts for the rest, $\Omega_h=1-\Omega_p$.
According to eq. (\ref{P4}), this solution exists only for
\be\label{LG9}
\alpha^2>\frac{n}{n-2}\ee
\be\label{LG10}
\alpha^2<\frac{n(6-n)}{3(4-n)(n-2)}\ee
which requires $n>3$.
At the upper bound (\ref{LG10}) for $\alpha^2$ the potential
energy $V$ goes to zero whereas at the lower bound (\ref{LG9})
the incoherent energy density $\rho$ vanishes and cosmology becomes
dominated by the homogenous scalar field or quintessence. The bounds
coincide for $n=3$ where the ``stability range'' for $\alpha$ shrinks
to the point $\alpha^2=3$. We note that for $\alpha$ in
the range $2<\alpha^2<3$ the
above solution is stable for $n=4$ but may shift to a quintessence
dominated universe as $n$ decreases below a critical value, for
example at matter-radiation equality. Such a behavior may lead to
an interesting interpretation of present cosmological observations.

Even though not all of the cases sketched here are realistic, it
becomes apparent that the presence of cosmon dark matter could have
an important backreaction effect on the evolution of homogenous
quintessence with nontrivial modifications of cosmology.

\section{Growth of cosmon fluctuations?}
\setcounter{equation}{0}

In this section we turn to the evolution of the large-scale cosmon
fluctuations. We adopt the particular exponential quintessence
potential discussed in the last section.
We have  seen previously that the cosmon fluctuations well within the
horizon behave as relastivistic matter. These short wavelength fluctuations
simply add to other forms of relativistic matter. Even if they
would dominate the energy density, they would behave similar
to standard radiation since $n=4$ and $\beta=0$. (For the
short-distance modes  the neglection of $\Delta\rho_k,\Delta p_k, \Delta
q_k$ (cf. eq. (\ref{XXC})) is justified.) In particular, the relativistic
cosmon fluctuations do not grow as compared to $\rho$.

A more interesting behavior may be expected for modes with wavelength
larger than, or comparable to, the horizon. There the gravity
effects become important.
We cannot neglect the retardation effects and have to solve
the coupled system of equations for the cosmon and gravity. With
$\tau=\ln t$ eq. (\ref{A112}) reads
\bear\label{x1}
&&\partial_\tau^2h_k+(3\eta-1)\partial_\tau h_k
+\left(\frac{\bar  
n}{6\alpha^2}(2+\zeta_0-6\eta^2\alpha^2)(1+4y_k)-\frac{\zeta_0}{\alpha^2}
\right)h_k+\frac{k^2t^2}{a^2}h_k\nonumber\\
&&\qquad=-\frac
{4}{M\alpha}(\zeta_0\delta\chi_k+\partial_\tau\delta\chi_k)\ear
and similarly for the scalar fluctuations eq. (\ref{LG1}) becomes
\be\label{x2}
\partial^2_\tau\delta\chi_k+(3\eta-1)\partial_\tau\delta\chi_k+\zeta_0\delta
\chi_k+\frac{k^2t^2}{a^2}\delta\chi_k=\frac{2M}{\alpha}
h_k(y_k+\eta+\eta y_k)\ee
For the modes far outside the horizon we can neglect the terms $\sim k^2
t^2/a^2$.
The resulting system of linear differential equations has
time-independent coefficients for the case of the quintessence
background discussed in the preceeding section. It can be
reduced to a system of first-order differential equations and
solved by standard methods.
The eigenvectors of the fluctuation matrix obey
\be\label{x3}
\delta\chi_k(\tau)=\delta\chi_k(0)e^{\lambda\tau},\quad  h_k(\tau)
=\frac{4s_0}{\alpha M}\delta\chi_k(\tau)\ee
where the  constants $\lambda$ and $s_0$ obey
\bear\label{x4}
&&s_0\lambda^2+(3\eta-1)s_0\lambda+xs_0+\zeta_0+\lambda=0\nonumber\\
&&\lambda^2+(3\eta-1)\lambda+\zeta_0-\frac{8}{\alpha^2}s_0w=0\ear
and
\be\label{7.6a}
x=\frac{\bar  
n}{6\alpha^2}(2+\zeta_0-6\eta^2\alpha^2)(1+4y_k)-\frac{\zeta_0}{\alpha^2}\quad  
, \quad w=y_k+\eta+\eta y_k\ee
With
\be\label{x5}
s_0=\frac{\alpha^2}{8w}(\lambda^2+(3\eta-1)\lambda+\zeta_0)\ee
the growth exponent $\lambda$ obeys the ``quartic''
equation\footnote{Note that $y_k$ and therefore $x$ and $w$ depend
on $\lambda$ such that eq. (\ref{x6}) is not simply a polynomial
in $\lambda$.}
\be\label{x6}
\lambda^4+2(3\eta-1)\lambda^3+\{\zeta_0+x+(3\eta-1)^2
\}\lambda^2+\{(3\eta-1)(\zeta_0+x)+\frac{8w}{\alpha^2}
\}\lambda+v=0\ee
with
\be\label{7.9a}
v=\zeta_0(x+\frac{8w}{\alpha^2})\ee
and we have to select the solution for $\lambda$ with the largest
real part. We note that eq. (\ref{x6}) has at least one positive
real solution if $v$ is negative.
A positive eigenvalue of $\lambda$ would imply that the large
scale fluctuations grow $\sim t^\lambda$. Such fluctuations would
be a good
candidate for becoming nonlinear.

In appendix A we have evaluated $y_k$ for the modes far outside the
horizon. We discuss possible solutions
where $h^i_0$ remains small for $k\to0$.
Then $y_k$ depends on $\lambda$ whereas it becomes
independent of $k$:
\be\label{NN1}
y=-(\lambda+2\eta)/(2\lambda+8\eta)\ee
For large $|\lambda|$ it approaches a constant, $y\to-1/2$,
whereas for $\lambda\to 0$ one finds $y\to -1/4$.
We have checked the validity of the assumption leading
to eq. (\ref{NN1}) for $\lambda=0$ (see next sect.).
(In any case, there is no reason that $y$ becomes unbounded
for $\lambda\to\infty$.) We conclude
that for large $\lambda$ the expression (\ref{x6}) grows $\sim\lambda^4$
whereas for $\lambda\to 0$ it approaches a constant
\be\label{NN2}
v_0=v(y(\lambda=0))=-\frac{\zeta_0}{\alpha^2}
(\zeta_0+2(1-3\eta))
=\frac{12\zeta_0\bar\beta}{\alpha^3\bar n}(\alpha^2\eta-1)\ee
which is negative for negative $\bar\beta$.
For $\bar\beta<0$ the existence of a positive eigenvalue is
therefore indeed guaranteed. For $\bar\beta=0$ there is at least
a constant mode $(\lambda=0)$ and there may be an
additional growing mode (cf. sect. 8).

Quite generally, the evolution of the modes far outside the horizon
is characterized by a nonvanishing ``particle production rate''
$\Delta_k$ in the sense of sect. 3. From
\be\label{x7}
\delta\dot\chi_k=\frac{\lambda}{t}\delta\chi_k\ee
one infers
\be\label{x8}
\beta_k=\frac{\lambda^2}{t^2}\alpha_k\ee
Inserting this behavior in eqs. (\ref{17}), (\ref{17a}), and
using $V''=\zeta_0t^{-2}$ (\ref{LG7}), we obtain
\be\label{x9}
\Delta_k=\frac{\zeta_0-\lambda^2}{\zeta_0+\lambda^2}n_k\ee
and therefore (cf. eq. (\ref{18a})) for the equation of state
\be\label{x10}
p_k=-\frac{\Delta_k}{n_k}(\rho_k-\Delta\rho_k)+\Delta  
p_k=\frac{\lambda^2-\zeta_0}{\lambda^2+\zeta_0}(\rho_k-
\Delta \rho_k)+\Delta p_k\ee

\section{Cosmon fluctuations in a radiation-dominated\protect\\ universe}
\setcounter{equation}{0}

Let us consider the evolution of the cosmon modes outside the
horizon in a radiation-dominated universe. We assume here that at
some early time the fluctuation amplitudes are small such that
the linear approximation applies. We want to know if the energy
density of the cosmon fluctuations can grow relative to the
radiation such that the cosmons finally dominate at large time.
The fluctuation
analysis of the preceeding sections
can easily be extended to this case. In particular,  the
evolution  equation (\ref{A111}) for $h$ remains unchanged. Fluctuations
in the  radiation
do not contribute to $\delta t^\mu_\mu$ due to $\delta p_r=
\delta\rho_r/3$.
We investigate here a possible solution where for $k\to 0$
the fluctuations $h^i_0$ can be neglected compared
to $h$ or $h_{00}$. In this case also the ratio $y=h_{00}/h$
becomes independent of additional radiation fluctuations in
the limit $k^2/(a^2H^2)\to 0$. (The estimate (\ref{A.22})
only involves the smallness of $h_{0i}$ as compared to $h$.
For the modes inside the horizon the ratio $y_k$ will depend
on an additional radiation component.) For a radiation-dominated
universe the background solution obeys
$\bar n=4, \bar\beta=0, \eta=1/2, \zeta_0=1$. This yields
\bear\label{NN3}
y&=&-\frac{1+\lambda}{4+2\lambda}\quad, \quad  
x=-\frac{2+3\lambda-\alpha^2\lambda}{\alpha^2(2+\lambda)},\nonumber\\
w&=&\ \frac{1-\lambda}{8+4\lambda} \quad,\quad  v_0=0\ear
and implies the existence of a solution
with stationary large-scale cosmon fluctuations, $\lambda=0$.
In addition, eq. (\ref{x6}) with eq. (\ref{NN1})
has a solution with a positive real $\lambda$ for $\alpha^2<2$. We
concentrate here on the case where the fluctuations far outside
the horizon $(k\to 0)$ are constant, i.e.
$\lambda=0$.

For the constant solution $(\lambda=0)$ we find
$(x=-1/\alpha^2\ ,\ w=1/8\ ,\ s_0=\alpha^2)$
\be\label{NN4}
h_k=\frac{4\alpha}{M}\delta\chi_k\ ,\quad (h_{00})_k=-\frac{\alpha}{M}
\delta\chi_k\ee
In eq. (\ref{x2}) the gravitational attraction precisely
compensates the driving force due to the potential.
It is easy to check with eq. (\ref{GG6}) that for this particular solution
the linear fluctuations in the energy momentum tensor vanish
\be\label{NN5}
\delta\rho_k=\delta p_k=0\ee
In consequence, the evolution equations for the linear metric
fluctuations (\ref{A.14}), (\ref{A.15}) have no source term
$\delta s_{00}=0\ ,\ \delta s^\mu_\mu=0$. This allows indeed a constant
solution for the metric provided $h_{00}=-h/4$ (cf. eq. (\ref{A.15})).
Furthermore, eq. (\ref{NN5}) gives a nice example that
nonvanishing cosmon fluctuations $\delta\chi_k$ lead not always
to energy density fluctuations in linear order. This is particularly
important for small positive $\lambda$ since by continuity
we may expect an approximate decoupling between the growth of $\delta\chi_k$
and $\delta\rho_k$. This may be relevant for reconciling comparatively
large cosmon fluctuations with very small fluctuations in the
radiation.

A simple picture for the cosmon fluctuations in the radiation-dominated
era emerges: For the modes outside the horizon the
fluctuations $\delta\chi_k$ and therefore also $\alpha_k$ remain constant.
Their contribution to the energy momentum tensor scales $\sim V$
and therefore decreases $\sim t^{-2}$, just as the energy density
in radiation and quintessence. On the other hand, the modes well inside
the horizon decrease in amplitude according to
\be\label{NN6}
\partial\ \ln\alpha_{k\gg k_p}/\partial\ \ln t=f_>=-1\ee
implying $\alpha_{k\gg k_p}\sim t^{-1}$. Their contribution
to the energy momentum tensor is dominated by the term $\sim
k^2\alpha_k/a^2$ (the contribution $\sim C/a^2$ in eq. (\ref{3.1}))
and therefore also decreases $\sim t^{-2}$. In short, the
cosmon dark matter scales proportional to the relativistic
matter, just as the homogenous quintessence contribution.

Using the relations (\ref{NN4}) we can also evaluate the
equation of state for the long wavelength fluctuations\footnote{Notice
that for $\lambda=0$ one has $\beta_k=0$ and therefore
$\Delta_k=n_k$.}. Since the linear contribution vanishes according
to eq. (\ref{NN5}), the lowest order contribution
is quadratic in the fluctuations.
We find that the mixed cosmon gravity contribution
$\Delta\rho, \Delta p$ dominates $(\rho_k=\rho_k^{(\delta\chi)}+
\Delta\rho_k$ etc.)
\bear\label{NN7}
\rho_k^{(\delta\chi)}&=&\frac{\zeta_0}{2t^2}\alpha_k\quad,\quad
p_k^{(\delta\chi)}\ =-\rho_k^{(\delta\chi)}\nonumber\\
(\Delta\rho)_k&=&-2\rho_k^{(\delta\chi)}\quad,\quad
\Delta p_k=2\rho_k^{(\delta\chi)}\nonumber\\
q^{(\delta\chi)}_k&=&\frac{\alpha\zeta_0}{2M}\frac{\alpha_k}{t^2}
\quad,\quad\Delta q_k\ =-2q^{(\delta\chi)}_k\ear
In particular, the large wavelength cosmon fluctuations give a negative
contribution\footnote{Here we use $\zeta_0=1$ as appropriate
for the radiation-dominated universe.} to the energy density!
\bear\label{NN8}
(\rho_c)_k&=&-(p_c)_k=-\frac{\alpha_k}{2t^2}\quad,\nonumber\\
(q_\varphi)_k&=&-\frac{\alpha}{M}\frac{\alpha_k}{2t^2}\ear
For the equation of state for the long-distance modes we find
$(p_c)_k/(\rho_c)_k=-1$ and the reader may  wonder how
this is compatible with a scaling $(\rho_c)_k\sim t^{-2}$. The
solution  of this puzzle arises from the term $q_\varphi$ in the cosmon
field equation. Indeed, the relation (\ref{2.10})
\be\label{NN9}
(\dot\rho_c)_k+3H((\rho_c)_k+(p_c)_k)+\dot\varphi(q_\varphi)_k=0\ee
is obeyed separately for every $k$-mode in the linear
approximation. For large $k$ we have a characteristic relativistic
equation of state $(p_c)_k=\frac{1}{3}(\rho_c)_k$
and $(q_\varphi)_k=0$ whereas for low $k$ the relations (\ref{NN8})
obey again (\ref{NN9}) for $(\rho_c)_k\sim t^{-2}$ due to
the nonvanishing $(q_\varphi)_k$. While for large $k$
the energy density $(\rho_c)_k\approx\rho_k^{(\delta\chi)}$ is
positive and decreases due to the expansion as usual, the negative
energy density for small $k$ increases as a result of an intricate
interplay between gravity and the scalar equation of motion! For
the convenience of the reader we summarize in table 1 the time
behavior of some relevant characteristic quantities for the
radiation-dominated era.

\bigskip
\begin{center}
\begin{tabular}{|c|c|}
\hline
small $k$& large $k$\\
\hline
$\delta\chi_k\sim const$& $\delta\chi_k\sim t^{-1/2}$\\
$\alpha_k\sim const$& $\alpha_k\sim t^{-1}$\\
$V''\sim t^{-2}$& $k^2/a^2\sim t^{-1}$\\
$\rho_k\sim t^{-2}$& $\rho_k\sim t^{-2}$\\
$p_k=-\rho_k$& $p_k=\frac{1}{3}\rho_k$
\\
\hline
\end{tabular}

\vspace{.5cm}
\noindent Table 1: Time dependence of characteristic quantities\\
in the radiation-dominated era

\end{center}

In addition, for the modes with small $k$ there is a relevant
pure gravitational contribution to the energy momentum tensor.
It is evaluated in appendix B. We find that $\rho_g$ and $p_g$
are of the same order as $\rho_c,p_c$ or smaller. We observe that nonzero
$\rho_g,p_g$ also decrease $\sim t^{-2}$. Since the cosmon energy
momentum tensor is separately conserved (cf. eq. (\ref{NN9})),
this must also hold for the pure gravitational energy momentum
tensor. In the pure gravitational sector there is no piece corresponding
to $q_\varphi$. Therefore the decrease $\rho_g\sim t^{-2}$ implies
that the pure gravitational part behaves similar to a relativistic
gas, $p_g=\rho_g/3$.

As a direct consequence of eq. (\ref{2.10}) and
$\rho_c\sim t^{-2}$ we can relate $q_\varphi=(\beta_c/M)\rho_c$ to
the equation of state for cosmon dark matter
\be\label{NN10}
\beta_c=\alpha\left(1-\frac{3\eta}{2}(1+\frac{p_c}{\rho_c})\right)
=\alpha(1-\frac{\eta}{2}n)\ee
In our case this is realized for $\eta=1/2$, yielding
\be\label{NN11}
\beta_c=\frac{\alpha}{4}(4-n)\ee
In dependence on the precise shape of the cosmon spectrum
$\beta_c$ varies between $0$ and $\alpha$ since $-\rho_c\leq
p_c\leq\rho_c
/3$. The relation (\ref{NN11}) differs from the result (\ref{3.20})
(as obtained for $\Delta_k=0$) by a factor of two and coincidence
occurs only for $p_c=\rho_c/3\ (n=4)$. This reflects the fact
that an unusual equation of state for cosmon dark matter is
related in our case to the effects of gravity for the
long distance modes for which $\Delta_k\not=0$. We emphasize, however,
that $\beta_c$ does not necessarily coincide with $\beta$ as defined
by eq. (\ref{27B}). Whenever the total energy density $\rho$ is
positive, the second equation in
eq. (\ref{NN8}) implies a negative value for $\beta$!

With $f_<=2\lambda$ and $\eta=1/2$ we can investigate
a cosmon spectrum of the type discussed in sect. 4. For the
solution with time-independent large-scale fluctuations
$(\lambda=0)$ it is
at best at the ``borderline of convergence'' (cf. eq. (\ref{S.16})).
Indeed, from eq. (\ref{S.15}) with $f_<=0$ we obtain
\be\label{NN12}
\nu_{IR}-\nu_{UV}=2\ee
For $\nu_{IR}=0,\ \nu_{UV}=-2$ the integral (\ref{18}) for the
energy density would diverge logarithmically both for small and large
$k$. On the other  hand, for a small positive $\nu_{IR}>0$ the spectrum
becomes infrared finite at the expense of a power divergence
in the ultraviolet. As discussed at the end of sect. 4, this
divergence may be cut off at a fixed scale $k_{UV}$
(which may be related to the physics of inflation). We may then
visualize cosmon dark matter as an effective mixture of two
components. The first ``radiation component'' accounts for
the modes well within the horizon, say $k>k_2(t),\ k_2(t)=100
k_p(t)$. If the UV-cutoff occurs when the
fluctuations remain still linear, the
``radiation component'' indeed behaves like relativistic matter to
high accuracy, being dominated by modes with $k\approx k_{UV}$.
The second ``horizon component'' accounts for the rest, whereby
the integrals for its contribution to the energy density etc.
are cut off at $k_2$. If the growth exponent for the modes
around the horizon (corresponding to $f_{max}$ in sect. 4) would be
negative or zero, $f_{max}\leq0$,
cosmon dark matter would  always be dominated by its radiation component.
However, our finding that the growth rate $f_<$ is zero for
the very long distance modes could well be compatible with a positive
$f_{max}$ as well. (The computation of $f_{max}$ is involved and
not done here.) A positive value of $f_{max}$ would have important
implications. Indeed, the relative weight of the horizon component
as compared to the radiation component scales\footnote{The factor
$k_p^{\ \nu_{IR}}$ (cf. eq. (\ref{S.3})) appears only if $k^3\rho_k$ develops
a maximum near $k_p$ due to the growth of the fluctuations at
this scale.} $\sim(k^2_p/k^2_{UV})(\alpha_{k_p}/\alpha_{k_{UV}})
k_p^{\nu_{IR}}$. It grows for a large enough growth rate of the modes
near the horizon, $f_H=\partial\ \ln\alpha_{k_p}/\partial\ \ln t$, namely
if $f_H$ obeys
\be\label{NN13}
f_H>f_>-(2+\nu_{IR})\frac{\partial\ \ln k_p}{\partial\ \ln t}
\quad, \quad f_H>2-4\eta+\nu_{IR}(1-\eta)\ee
For the radiation-dominated era this requires $f_H>\nu_{IR}/2$.
The energy density in the horizon component of cosmon
dark matter would then decrease only as
\be\label{8.11a}
\rho_{c,H}\sim t^{2\lambda_H-\nu_{IR}-2}\ee
Since the homogenous
radiation  scales
$\sim t^{-2}$, a positive value of $f_H-\nu_{IR}/2$ would  lead
to a situation where the horizon component of cosmon dark matter
finally not only
overwhelms the radiation component of cosmon dark matter, but the
total energy density in radiation as well! When this happens, the
radiation-dominated era would end and transite to an era
dominated by the horizon component of cosmon dark matter!

We emphasize that due to the negative value (\ref{NN8}) of
$q_\varphi$ the value of $\bar\beta$ is actually negative in
a radiation-dominated universe. This effect is tiny
as long as radiation dominates substantially. Once the horizon
component of cosmon dark matter becomes more important,
however, the size of $|\bar\beta|$ increases. In turn, this leads
to a larger value of $\lambda$ (cf. eq. (\ref{NN2})) and
further increases the growth of the cosmon fluctuations
around and beyond the horizon! We also recall
at this point that quintessence with an exponential potential may
be an oversimplified example. It is well conceivable \cite{CW1} that the
effective value of $\alpha$ depends on $\varphi$ and therefore
on time. If $\alpha$ decreases with time, an increase of the horizon
component of cosmon dark matter becomes even more plausible.

In conclusion, we have sketched here
a first plausible scenario where cosmon dark matter grows
relative to radiation. In this scenario the cosmon spectrum
has the qualitative shape (\ref{S.1}) with a finite energy
density. Cosmon dark matter will finally overwhelm the radiation
and bring the radiation-dominated epoch to an end. Also the cosmon
fluctuations grow nonlinear during this evolution.
Equally interesting, but more difficult to treat, is the
case where the effective UV- (for $\nu_{IR}>0$)
or $IR$- (for $\nu_{IR}<0$) cutoff arises as a consequence
of nonlinearities. The effective equation of state for cosmon dark
matter requires then a quantitative understanding of nonlinear
fluctuations.

\section{Cosmon-dominated universe}
\setcounter{equation}{0}

In this section we briefly discuss  a universe
where homogenous quintessence and
the cosmon fluctuations dominate the energy momentum
tensor. Such a scenario could be realized in the ``recent'' past
of our universe (say for $z<(10^2-10^3)$). The different
relevant components of such a universe are gravity,
homogenous quintessence,
(the energy density of the scalar background field) and cosmon
dark matter (the incoherent part of the expectation value of the
scalar energy density). In principle, there are two
possibilities: either the scaling law of cosmon dark
matter is similar to the usual behavior with $\rho\sim t^{-2}$,
or even the scaling law is modified. For the first case we may
look for a
solution of the type discussed in sect. 6. In this case we
have to determine self-consistently
\bear\label{NN15}
\bar n&=&n=3(1+(p_c+p_g)/(\rho_c+\rho_g))\nonumber\\
\bar\beta&=&\beta=Mq_\varphi/(\rho_c+\rho_g)\ear
We require $\rho=\rho_c+\rho_g>0$ and infer
from eq. (\ref{NN8}) that $\bar\beta$ is different from
zero and negative.

The computation of $\bar n$ and $\bar\beta$ is rather
involved and should be done in a self-consistent way. First one has
to evaluate $p,\rho$ and $q_\varphi$ by performing the fluctuation
analysis for a family of background solutions characterized by
$\bar\beta$ and $\bar n$. Then one has to solve the ``self-consistency
equation'' (\ref{NN15}) for $\bar\beta$ and $\bar n$. The solution
will correspond to $\bar\beta<0$ and we also expect $\bar n<4$. In view
of the linear fluctuation analysis of sec. 7 the existence
of such a solution with $\lambda_H>0$ seems not unlikely.
In this respect the precise shape of the cosmon spectrum
will matter. Indeed, the relation for the growth rate of the cosmon
energy density
\be\label{NN16}
\partial\ \ln\rho_c/\partial\ \ln t=\partial\ \ln A/\partial
\ \ln t-2=2\lambda_H-(1-\eta)\nu_{IR}-2=-2\ee
implies
\be\label{NN17}
\lambda_H=(1-\eta)\nu_{IR}/2\ee

This scenario is not the only possibility and the
time evolution of the various components of the energy
denstiy can have a much richer structure. In the remainder
of this section we summarize a few generic
features of a cosmon-dominated universe. This
discussion applies quite generally, far beyond
the specific models discussed in this paper. In presence of
quintessence and
substantial cosmon fluctuations the composition of the
energy momentum tensor becomes fairly complex.
For a cosmon-dominated universe
one may approximate  $\rho_{tot}=
\rho_h+\rho_{inhom}=\rho_h+\rho_c+\rho_g$. The general
characteristics of a cosmon-dominated universe can be
summarized as follows:\\
(i) The total energy momentum tensor being conserved, i.e.
\be\label{9.4}
\dot\rho_{tot}+3H(\rho_{tot}+p_{tot})=0\ee
we can infer from eq. (\ref{2.12}) the ``rate of exchange''
between homogenous and inhomogenous quintessence. (We use
$\dot\varphi\geq0$.)
\bear\label{9.5}
&&\dot\rho_h+3H(\rho_h+p_h)=q_\varphi\dot\varphi=\frac{\beta}{M}
\rho_{inhom}\sqrt{\rho_h+p_h}\nonumber\\
&&\dot\rho_{inhom}+3H(\rho_{inhom}+p_{inhom})
=-q_\varphi\dot\varphi=-\frac{\beta}{M}\rho_{inhom}
\sqrt{\rho_h+p_h}\ear
It is proportional to $\beta$.\\
(ii) The expansion of the universe accelerates if
\be\label{9.6}
\frac{\ddot a}{a}=\dot H+H^2=-\frac{1}{12M^2}(\rho_{tot}
+3p_{tot})>0\ee
which is equivalent to a total equation of state  
$w_{tot}=p_{tot}/\rho_{tot}<-1/3$. In terms of the
equation of state for homogenous and inhomogenous quintessence
$p_h=w_h\rho_h$, $p_{inhom}=w_{inhom}\rho_{inhom}$
we can express
\be\label{9.7}
w_{tot}=w_h\Omega_h+w_{inhom}(1-\Omega_h)=w_h\Omega_h
+w_c\Omega_c+w_g\Omega_g\ee
(iii) If gradient terms dominate over the time derivatives
the cosmon equation of state is negative, $w_c\leq-1/3$
\cite{CWBR}. This may, however, be (partially) compensated
by $w_g\Omega_g>0$. \\
(iv) For fixed $\beta$ and $w_h$ we may write
\be\label{9.8}
\dot\rho_{inhom}+n^{eff}_{inhom}H\rho_{inhom}=0\ ,\quad
n^{eff}_{inhom}=3(1+w_{inhom})+\beta\sqrt{6\Omega_h(1+w_h)}
\ee
In case of negative $\beta$ and $w_{inhom}$ one finds
$n^{eff}_{inhom}<3$. This would imply that the cosmon energy density
dilutes shower than non-relativistic  matter and, a fortiori,
much slower than radiation.\\
(v) A decrease of $\beta$ towards negative values slows down the
decrease of $\varphi$ (cf. eq. (\ref{2.12}) with $q_\varphi<0$).
This effect reduces the kinetic energy of the background field as
compared to its potential energy and therefore drives $w_h$
towards negative values. Details depend on the form of the
cosmon potential.\\
(vi) The relative change in homogenous quintessence obeys
\be\label{9.9}
\dot\Omega_h=H(1-\Omega_h)\{3\Omega_h(w_{inhom}-w_h)
-\beta\sqrt{6(1+w_h)}\}\ee
Negative $\beta$ and negative $w_h$ tend to increase
$\Omega_h$, whereas negative $w_{inhom}$ has the
opposite effect.\\
(vii) It is conceivable that a change in the effective
value of $\beta$ and $\omega_{inhom}$ after structure formation
trigger an increase of $\Omega_h$ and an accelerated
expansion of the universe \cite{CWBR}.

We postpone a more detailed discussion of the interesting cosmon-dominated
universe to a future investigation. Here we hope that we have motivated
that such an investigation is worthwhile. We should be ready to encounter
surprising effects as compared to standard cold dark matter
cosmology with homogenous quintessence: The equation of state of cosmon dark
matter may neither be the one of a relativistic nor of a nonrelativistic
gas. It could even change with time. Homogenous quintessence has most likely
a substantial coupling to cosmon dark matter, even though its
coupling to ordinary matter vanishes or is very small. This introduces
features that are not familiar in the standard hydrodynamical
treatment of dark matter. In particular, the conservation
equation (\ref{2.10}) contains a modification $\sim q_\varphi$.

In order to get some insight into the effects of
cosmon dark matter, for example by numerical simulations, one
may first treat $\beta$ and $w_{inhom}$ as free parameters.
Even though this is only a parametrization of our lack of
knowledge we may learn how structure formation and the cosmic
microwave background (CMB) are modified as compared to a standard
cold dark matter inhomogenous component (for which $\beta=w_{inhom}
=0$). We note that for cosmon dark matter $w_{inhom}$ could even
be negative \cite{CWBR}.

We also note that large cosmon fluctuations are not
Gaussian any more, despite the likely approximate
Gaussian distribution in early cosmology when the fluctuations
were in the linear regime. The effect of the non-Gaussian
behavior on the time evolution of the cosmon fluctuations will
be briefly addressed in the next section. We recall, however,
that even strong non-Gaussian cosmon fluctuations are compatible
with approximately Gaussian fluctuations in the cosmon energy
density and the CMB.

\section{Large fluctuations in mean field approximation}
\setcounter{equation}{0}

In a cosmon-dominated universe the cosmon fluctuations are not small
anymore and linear analysis does not apply. One may encounter
gravitationally bound cosmon lumps \cite{CL} on all length scales.
Extended cosmon lumps may become the seeds for galaxies and could describe
the observed galactic halos. In this section we make a modest
attempt to get a glance on the modifications of the cosmon
fluctuation equations which are induced by the nonlinearities.
We study a mean field approximation. This takes into account
part of the effects of nonlinearities, while keeping a structure
of the fluctuation equations similar to the linear analysis. In
particular, the effective potential and the effective cosmon mass
will depend on the amplitude of the fluctuations. Typically, the mass
increases for the nonlinear system such that the inverse mass
(correlation length) may become substantially smaller than the
horizon. We notice, however, that the mean field analysis will
fail to give a reliable description once gravitationally bound cosmon
lumps form.

As a general framework for the analysis of nonlinear effects we
may use the time-dependent effective action \cite{NE}. For
simplicity we neglect here the gravitational effects. This is only
justified if nonlinearities occur also at scales somewhat smaller
than the horizon. Since this is indeed the tendency of the mean
field approximation we can investigate such a scenario in a
self-consistent way. We start with the microscopic
equation for the cosmon fluctuation
\be\label{5.1}
\delta\ddot\chi_k+3H\delta\dot\chi_k+\frac{k^2}{a^2}\delta\chi_k=f_k
[\delta\chi]
=-\int d^3 xe^{-ikx}(V'(\varphi+\delta\chi)-V'(\varphi))\ee
The quantity $f_k$ depends on the details of the potential.
For our  example of an exponential potential (\ref{LG6}) $f_k$ can be
represented by a Taylor series
\be\label{CY}
f_k=\frac{\alpha}{M}V(\varphi)\int d^3xe^{-ikx}\sum_{n=1}^\infty
\frac{1}{n!}(-\frac{\alpha}{M}\delta\chi(x))^n\ee
We introduce $\delta\pi_k=\delta\dot\chi_k$ and
write eq. (\ref{5.1}) as a coupled system of first-order
differential equations (with $\delta\chi^n_k$ the Fourier
transform of $\delta\chi^n(x)$)
\bear\label{DY}
\delta\dot\chi_k&=&\delta\pi_k\nonumber\\
\delta\dot\pi_k&=&-3H\delta\pi_k-\frac{k^2}{a^2}\delta\chi_k+\frac{\alpha}
{M}V(\varphi)\sum^\infty_{n=1}\frac{1}{n!}(-\frac{\alpha}{M}
\delta\chi)^n_k\ear
For a given time evolution of the background quantities
$a(t), H(t)$ and $\varphi(t)$ this is a closed system of
``microscopic'' equations. The time evolution of correlation functions
for such systems can be described in terms of an exact
evolution equation for a time-dependent effective action \cite{NE}.
We will be satisfied here with a simple ``mean field''
or ``Hartree''-type approximation. We approximate
\be\label{EY}
\delta\chi_k^n=n<\delta\chi^{n-1}>\delta\chi_k\ee
with $<\delta\chi>=0$ and ($m$ integer, $m\geq2$)
\bear\label{FY}
&&<\chi^{2m-1}>=0\nonumber\\
&&<\chi^{2m}>=3\cdot5...(2m-3)(2m-1)(<\chi^2>)^m\ear
The expectation value of the squared fluctuations is given
again by eq. (\ref{3.2}), $<\delta\chi^2>=A$. We note that $A$ receives
contributions from all momentum modes, both larger and smaller
than $k$. Already in the mean field approximation the evolution of
$\delta\chi_k$ is therefore coupled to all other momentum modes.

In terms of $\alpha_k,\beta_k$ (cf. eq. (\ref{3.3}))
and $\gamma_k$ (cf. eq. (\ref{3.12})) one obtains a closed
system for the Fourier components of the two-point correlation
\bear\label{GY}
&&\dot\alpha_k\equiv\beta_{\alpha_k}=2\gamma_k\nonumber\\
&&\dot\gamma_k\equiv\beta_{\gamma_k}=\beta_k-3H\gamma_k
-\left(\frac{k^2}{a^2}+\mu^2
\right)\alpha_k\nonumber\\
&&\dot\beta_k\equiv\beta_{\beta_k}
=-6H\beta_k-2\left(\frac{k^2}{a^2}+\mu^2\right)\gamma_k\ear
From
eq. (\ref{GY})  we can derive the useful identity
\[\partial_t(\alpha_k\beta_k-\gamma_k^2)=-6H(\alpha_k\beta_k-\gamma_k^2)\]
One infers a separate conservation law for every momentum mode
\[a^6((\alpha_k\beta_k-\gamma_k^2)=const\]
A modified effective mass term $\mu^2$ governs the cosmon
fluctuations
\be\label{HY}
\mu^2=\frac{\alpha^2}{M^2}V\left(1+\sum^\infty_{m=1}g_m\left(\frac{\alpha
^2A}{M^2}\right)^m\right)=\frac{\alpha^2}{M^2} V{\cal G}\left(\frac{
\alpha^2A}{M^2}\right)\ee
with
\be\label{IY}
g^{-1}_m=2\cdot4\cdot6\cdot8\cdots(2m-2)(2m)\ee
The new ingredient as compared to the linear approximation
is the function ${\cal G}\left(\frac{\alpha^2A}{M^2}\right)$
which is equal to one for small fluctuations. We conclude that
the mean field approximation leads to a time-dependent mass
term which depends itself on the size of the fluctuations.

The function ${\cal G}$ also appears in the nonlinear
expressions for $\rho_c,p_c$ and $q_\varphi$
\bear\label{JY}
&&\rho_c=V(\varphi)[{\cal G}\left(\frac{\alpha^2A}{M^2}\right)
-1]+\frac{1}{2}B+\frac{C}{2a^2}\nonumber\\
&&p_c=-V(\varphi)[{\cal G}\left(\frac{\alpha^2A}{M^2}\right)
-1]+\frac{1}{2}B-\frac{C}{6a^2}\nonumber\\
&&q_\varphi=\frac{\alpha}{M}V(\varphi)[{\cal G}\left(\frac{\alpha^2
A}{M^2}\right)-1]\ear
and we may derive the relations
\be\label{KY}
{\cal G}\left(\frac{\alpha^2A}{M^2}\right)=1+\frac{2B-\rho-3p}{
2V(\varphi)}\ee
\be\label{LY}
\mu^2=\frac{\alpha^2}{M^2}(V+B-\frac{\rho}{2}-\frac{3p}{2})\ee
Comparison with the linearized expressions (\ref{3.1}) shows that
the nonlinearities result essentially in a substitution
$V''(\varphi)A/2\to\mu^2M^2/\alpha^2-V$, or
\bear\label{MY}
&&\rho_c=\frac{\mu^2M^2}{\alpha^2}-V+\frac{1}{2}B+\frac{C}
{2a^2}\nonumber\\
&&p_c=-\left(\frac{\mu^2M^2}{\alpha^2}-V\right)+\frac{1}{2}B
-\frac{C}{6a^2}\nonumber\\
&&q_\varphi=\frac{\alpha}{M}\left(\frac{\mu^2M^2}{\alpha^2}-V\right)
\ear
One may define a further effective mass term (different from $\mu^2$!)
\be\label{39a}
m^2_\rho=2V({\cal G}-1)/A=\frac{2}{A}\left(\frac{\mu^2M^2}{\alpha^2}
-V\right)=\frac{2M^2}{\alpha^2A}\frac{{\cal G}-1}{\cal G}\mu^2\ee
which replaces $V''$
for the scalar contribution to $\rho_c$ etc. The nonlinear
effects are reflected in the difference between $\mu^2$, $m^2_\rho$
and $V''(\varphi)$,
\bear\label{NY}
&&\frac{\mu^2-V''(\varphi)}{V''(\varphi)}=\sum^\infty_{m=1}g_m\left(\frac
{\alpha^2A}{M^2}\right)^m,\nonumber\\
&&\frac{m^2_\rho-V''(\varphi)}{V''(\varphi)}=\sum^\infty_{m=1}
2g_{m+1}\left(\frac
{\alpha^2A}{M^2}\right)^m\ear

We conclude that the
discussion of the cosmon fluctuations in the mean field
approximation parallels the linear approximation in many
aspects, with two important modifications:

(a) the mass term $\mu^2$ replaces $V''$ in eqs. (\ref{17}),
(\ref{17a}), \ref{19}), (\ref{22a}), (\ref{3.13}),

(b) the relation (\ref{3.1}) and similar relations for $\rho_k$ and
$p_k$ get modified by replacing $V''$ by $m^2_\rho$.

Let us
first consider the short distance modes with $k^2/a^2\gg\mu^2,H^2$. We
assume that $A$ varies only slowly  on the time scale of the
corresponding oscillation and neglect
$\dot A$ in a first approximation. One therefore finds again in leading
order that particle production is suppressed, $\Delta_k=0$. The relation
between $\rho_k,p_k$ and $n_k$ reads
now (for $\Delta_k=0$ and the definition of $n_k$ involving $\mu^2$
instead of $V''$)

\bear\label{N1}
\rho_k&=&\frac{1}{2}\left\{\left(\frac{k^2}{a^2}+m^2_\rho\right)
\alpha_k+\beta_k\right\}=\left(\frac{k^2}{a^2}+\mu^2\right)^{1/2}n_k+
\frac{1}{2}\Delta m^2\alpha_k\nonumber\\
p_k&=&\frac{1}{2}\left\{\beta_k-\left(m_\rho^2+\frac{k^2}{3a^2}
\right)\alpha_k\right\}=\frac{k^2}{3a^2}\left(\frac{k^2}{a^2}
+\mu^2\right)^{-1/2}n_k-\frac{1}{2}\Delta m^2 \alpha_k\ear
Beyond the replacement of $V''$ by $\mu^2$ we observe additional
terms involving the difference of the two types of effective masses

\be\label{N2}
\Delta m^2=m^2_\rho-\mu^2=-V''\sum^\infty_{m=1}\frac{m}
{m+1}g_m\left(\frac{\alpha^2A}{M^2}\right)^m\ee
For large enough $k^2/a^2$ this new effect is
suppressed, however, yielding a typical relative correction
to the equation of state $\sim\Delta m^2a^2/k^2$. We
conclude that in the mean field approximation the high momentum modes
again behave as relativistic particles. If they would
dominate, the equation of state would be
$p\approx\rho/3$. Neglecting the time evolution of $a$,
$V$ and $A$, the deviation from the relativistic behavior due to masslike
terms is governed by
\be\label{N3}
\frac{p_k}{\rho_k}=\frac{1}{3}\left[1+\frac{a^2}{k^2}(\mu^2-2m^2_\rho)+
0\left(\left(\frac{m^2a^2}{k^2}\right)^2\right)\right]\ee
In lowest order in $\alpha^2A/M^2$ the nonlinear effects
cancel for this relation, $\mu^2=-2m^2_\rho=-V''(1+0((
\alpha^2A/M^2)^2)$.

We conclude that in the mean field approximation
the short distance modes $(k^2/a^2\gg\mu^2)$ behave
as relativistic matter even in presence of nonlinear effects.
In this approximation the interesting candidates for cosmon dark matter are  
therefore fluctuations
with $k^2/a^2$ comparable to $\mu^2,m^2_\rho$ or smaller.
An interesting situation would arise if
these fluctuations have a wavelength well within the horizon, which
requires $\mu^2\gg H^2$ or $m^2_\rho\gg H^2$. In order to see if
this is possible, we first note the inequalities
\be\label{N4}
\rho_c\geq \frac{1}{2} m^2_\rho A\ ,\quad H^2\geq \frac{1}{12}
\frac{m^2_\rho A}{M^2}
\ee
On the other hand, nonlinear effects are important only for
$\frac{A}{M^2}\geq\frac{1}{\alpha^2}$, and we conclude
for the nonlinear regime
\be\label{N5}
\frac{m^2_\rho}{H^2}\leq 12\alpha^2\ee
For $\alpha$ sufficiently large $m_\rho$ could indeed be substantially
larger than $H$. Furthermore, for large $x=\alpha^2A/M^2$ the ratio
$\mu^2/m^2_\rho\approx x/2$ is also large.

\section{Conclusions}

In conclusion, the cosmon dark matter scenario seems plausible
enough to merit a detailed investigation. Several important questions
have been left open in the present paper: a precise treatment of the
time evolution of fluctuations with wavelength around the horizon,
a quantitative investigation of a possible cosmon-dominated epoch,
a discussion of the time evolution of fluctuations in the radiation
(relevant for the CMB), a description of structure
formation, the role of possible gravitationally bound cosmon
lumps, the influence of structure formation on the time evolution
of quintessence... The present note presents only the
basic ideas and first arguments in favor of a cosmon dark matter
scenario. In particular, we have discussed a setting where the energy
density of the cosmon fluctuations grows relative to radiation during
a radiation-dominated period. Our paper
also develops a general theoretical framework to
investigate the relevant questions in a context where nonlinearities
and backreaction effects become important. Much remains to be done,
however, before a consistent ``cosmon dark matter cosmology'' is developed.

Nevertheless, two important general conclusions can be drawn for a possible
cosmon dark matter scenario. (1) An unusual equation of state may
arise not only for the homogenous quintessence component but
also for the part of dark matter taking part in structure formation.
The effective equation of state may depend on time and perhaps
even on the particular problem. It is conceivable that the effective
equation of state differs for extended spherically symmetric
objects (e. g. cosmon lumps) and for a homogenous cosmology. (2)
One expects a sizeable coupling of quintessence to cosmon dark matter.
This modifies the energy momentum conservation for dark matter and
influences structure formation as well as the details of the
cosmic microwave background.

Most important, the cosmon dark matter scenario opens the perspective
of a unified description of homogenous dark energy and ``clumping''
dark matter. It has the potential to explain the rough equality
of these two components as well as to answer the question of
``why now'' for the present acceleration, without putting in
the answers ``by hand'' in the form of
appropriately tuned potentials or kinetic terms. A success of
these ideas would
greatly enhance the naturalness of quintessence. For experimental
dark matter searches the consequences would be dramatic: cosmons
interact only with gravitational strength and cannot be detected
by present and planned laboratory experiments. No new dark matter
particle would be needed. In the forseeable future dark matter would
show up only in an astrophysical or cosmological context!

\section*{Acknowledgement}
The author would like to thank C. M. M\"uller for pointing
out an omission in a previous formulation of appendix A.

\section*{Appendix A\protect\\
Evolution equation for metric fluctuations}
\renewcommand{\theequation}{A.\arabic{equation}}
\setcounter{equation}{0}

In this appendix we derive the microscopic field equation for the
fluctuations of the metric, $h_{\mu\nu}=g_{\mu\nu}-\bar g_{\mu\nu}$, in
linear order. We concentrate on a Robertson-Walker (zero curvature)
background  $\bar g_{\mu\nu}$ and the harmonic background gauge
$D_\mu h^\mu_{\ \nu}=\frac{1}{2}
\partial_\nu h$. The starting point is the Einstein equation
\be\label{A.1}
R_{\mu\nu}-\frac{1}{2}Rg_{\mu\nu}=\frac{1}{2M^2}t_{\mu\nu}\ee
The linear expansion for the Fourier modes with comoving
momenta $k\not=0$ yields
\be\label{A.2}
(\delta R_{\mu\nu})_k=\frac{1}{2M^2}(\delta s_{\mu\nu})_k\ee
with
\bear\label{A.2a}
s_{\mu\nu}&=&t_{\mu\nu}-\frac{1}{2}t^\rho_\rho g_{\mu\nu}\quad,\quad
\delta s_{\mu\nu}=s_{\mu\nu}-\bar s_{\mu\nu},\nonumber \\
\bar s_{\mu\nu}&=&\hat T_{\mu\nu}-\frac{1}{2}\hat T_{\rho\sigma}
\bar g^{\rho\sigma}\bar g_{\mu\nu}\ear
We observe that we have to subtract the background energy momentum
tensor which includes fluctuation effects that are already taken
into account in the determination of the background solution. Therefore
$\delta t_{\mu\nu}=t_{\mu\nu}-\hat T_{\mu\nu}$ is the microscopic
energy tensor for the cosmon and metric fluctuations around the
background solution. Here
\be\label{A.2b}
\hat T_{\mu\nu}=T^\varphi_{\mu\nu}+T_{\mu\nu}\quad,\quad
T^\varphi_{\mu\nu}=\partial_\mu\varphi\partial_\nu
\varphi+(\frac{1}{2}\dot\varphi^2-V)\bar g_{\mu\nu}\ee
includes the incoherent effects in $T_{\mu\nu}$. As a result,
the homogenous part ($k=0$ component) has always to be subtracted.
Defining $V(\varphi+\delta\chi)=V(\varphi)+\delta V$ one has,
for example, the fluctuation contribution
\be\label{A.2c}
(V(\varphi+\delta\chi)-<V>)_k=\delta V_k-\delta V_{k=0}\ee

We first need the expansion of the Ricci tensor
\bear\label{A.3}
(\delta R_{\mu\nu})_k&=&(\partial_\lambda\delta\Gamma_{\mu\nu}^{\ \  
\lambda})_k-(\partial_\mu\delta\Gamma_{\lambda\nu}^{\ \ \lambda})_k\nonumber\\
&&+\Gamma_{\mu\nu}^{\ \ \tau}(\delta\Gamma_{\lambda\tau}^{\ \ \lambda}
)_k+\Gamma_{\lambda\tau}^{\ \ \lambda}(\delta\Gamma_{\mu\nu}^{\ \ \tau})_k
-\Gamma_{\lambda\nu}^{\ \ \tau}(\delta\Gamma_{\mu\tau}^{\ \ \lambda})_k
-\Gamma_{\mu\tau}^{\ \ \lambda}(\delta\Gamma_{\lambda\nu}^{\ \ \tau})_k\ear
Here $\Gamma_{\mu\nu}^{\ \ \lambda}$ is the connection for the
homogenous and isotropic background metric and
\bear\label{A.4}
(\delta\Gamma_{\mu\nu}^{\ \ \lambda})_k&=&\frac{1}{2}\bar g^{\lambda\rho}
(\partial_\mu h_{\nu\rho}+\partial_\nu
h_{\mu\rho}-\partial_\rho h_{\mu\nu})_k- \Gamma_{\mu\nu}^{\ \ \tau}(h_\tau
^{\ \lambda})_k\nonumber\\
&=&\frac{1}{2}(h_{\nu\ ;\mu}^{\ \lambda}+h_{\mu\ ;\nu}^{\ \lambda}
-h_{\mu\nu;}^{\ \ \ \lambda})\ear
We omit in the following the bar for the background
metric which is used to raise and lower the indices of
$h_{\mu\nu}$ and also the momentum label $k$ for $h_{\mu\nu}$.
One obtains $(h=h_\lambda^{\ \lambda})$
\bear\label{A.5}
\delta R_{\mu\nu}&=&(\delta\Gamma_{\mu\nu}^{\ \ \lambda})_{;\lambda}-(
\delta\Gamma_{\lambda\nu}^{\ \ \lambda})_{;\mu}\nonumber\\
&=&\frac{1}{2}(h_{\mu\ ;\nu\lambda}^{\ \lambda}+h_{\nu\ ;\mu\lambda}^{\  
\lambda}-h_{\mu \nu;\ \lambda}^{\ \ \ \lambda}-h_{;\nu\mu})\ear
We note that with respect to gauge transformations of the background
metric $h_{\mu\nu}$ transforms as a tensor. Using the identity
\be\label{A.6}
h_{\mu\  ;\nu\tau}^{\ \lambda}-h_{\mu\  ;\tau\nu}^{\ \lambda}
=R_{\nu\tau\ \mu}^{\ \ \sigma}h_\sigma^{\ \lambda}-R_{\nu\tau\ \sigma}
^{\ \ \lambda}h_\mu^{\ \sigma}\ee
this yields
\be\label{A.7}
\delta R_{\mu\nu}=-\frac{1}{2}h_{\mu\nu;\ \lambda}^{\ \ \  \lambda}
-R_{\mu\lambda\nu\sigma}h^{\lambda\sigma}
+\frac{1}{2}R_{\nu\sigma}h_\mu^{\ \sigma}+\frac{1}{2}R_{\mu\sigma}h_\nu^{\  
\sigma}+\delta K_{\mu\nu}\ee
with
\be\label{A.8}
\delta K_{\mu\nu}=\frac{1}{2}(h_{\mu\ ;\lambda\nu}^{\ \lambda}-\frac{1}{2}
h_{;\mu\nu}+h^{\ \lambda}_{\nu\ ;\lambda\mu}-\frac{1}{2}
h_{;\nu\mu})\ee
The harmonic background gauge implies $\delta K_{\mu\nu}=0$. In particular,
one finds for the trace
\be\label{A.9}
g^{\mu\nu}\delta R_{\mu\nu}=-\frac{1}{2}h_{;\ \mu}^{\ \mu}\ee
which yields the field equation for the trace of the metric fluctuations in
an arbitrary background
\bear\label{A.10}
h_{;\ \mu}^{\ \mu}&=&-\frac{1}{M^2}\delta s^\mu_{\mu}=\frac{1}{M^2}
(\delta t^\mu_\mu-2\hat T^{\mu\nu}h_{\mu\nu}+\frac{1}{2}
\hat T^\rho_\rho h^\mu_\mu)\nonumber\\
&=&\frac{1}{M^2}[\delta t^\mu_\mu-\frac{1}{2}(\dot\varphi^2+\rho+p)(h+4h_{00})]\ear
where $\delta t^\mu_\mu=\bar g^{\mu\nu}\delta  
t_{\mu\nu}\not=\delta(g^{\mu\nu}t_{\mu\nu})$.

We also need an evolution equation for $h_{00}$, i.e.
\be\label{A.11}
h_{00;\ \lambda}^{\ \ \ \lambda}+2R_{0\lambda 0\sigma}h^{\lambda\sigma}
-2R_0^{\ \sigma}h_{0\sigma}=-\frac{1}{M^2}
\delta s_{00}\ee
Inserting the connection for the Robertson-Walker metric,
$\Gamma_{ij}^{\ \ 0}
=Hg_{ij},\Gamma_{0i}^{\ \ j}=H\delta_i^j$, zero otherwise,
the curvature tensor reads $R_{0i0}^{\ \ \ j}=-
(H^2+\dot H)\delta^j_i,\ R_{00}=-3(H^2+\dot H)$ such that
\be\label{A.12}
h_{00;\ \lambda}^{\ \ \ \lambda}-2(H^2+\dot H)(h+4h_{00})=
-\frac{1}{M^2}\delta s_{00}\ee
We finally employ the explicit expressions $h_{;\ \lambda}
^{\ \lambda}=-(\ddot h
+3H\dot h+(k^2/a^2)h),h_{00;\ \lambda}^{\ \ \ \lambda}
=-(\ddot h_{00}+3H\dot h_{00}+(k^2/a^2)h_{00}
-2H^2(h+4 h_{00})+4H\partial_i h_0^i)$ and the gauge condition
\be\label{A.13}
\partial_ih_0^i=\frac{1}{2}\dot h+\dot h_{00}+H(h+4h_{00})\ee
and infer
\bear
&&\ddot h+3H\dot h+\frac{k^2}{a^2}h=\frac{1}{M^2}\delta s^\mu_\mu\label
{A.14}\\
&&\ddot h_{00}+H(2\dot h+7\dot h_{00})+2(2H^2+\dot H)(h+4h_{00})
+\frac{k^2}{a^2}h_{00}=\frac{1}{M^2}\delta s_{00}\label{A.15}
\ear
Using again the harmonic gauge condition
\be\label{A.16}
\partial_j h_i^{\ j}=\frac{1}{2}\partial_i h+\dot h_{0i}+3H h_{0i}\ee
we obtain similarly
\be\label{A.17}
\ddot h_{0i}+3H\dot h_{0i}+\frac{k^2}{a^2}
h_{0i}+(6H^2+5\dot H)h_{0i}
+H(\partial_i h+2\partial_i h_{00})
=\frac{1}{M^2}\ \delta s_{0i}\ee

For a solution of the metric field equation we need
the source terms $\delta s_{\mu\nu}$. Defining
$\delta t_{00}=\delta\rho,\ \delta t_{ij}=\delta p\bar g_{ij}$
one obtains
\be\label{GG3}
\delta s_{00}=\frac{1}{2}[\delta\rho+3\delta p-(p-V+\frac{1}{2}\dot
\varphi^2)(h+4h_{00})]\ ,\ \bar g^{ij}
\delta s_{ij}=\frac{3}{2}(\delta\rho-\delta p)+\frac{1}{2}(\rho+V+\frac{1}
{2}\dot\varphi^2)(h+4h_{00})\ee
or
\be\label{GG4}
\delta s^\rho_\rho=\delta\rho-3\delta p+\frac{1}{2}
(\dot\varphi^2+\rho+p)(h+4h_{00})\ee
For the cosmon the microscopic fluctuations $\delta \rho$ and $\delta p$
are linear in the fluctuations $\delta\chi, \delta\dot\chi$,
in contrast to usual matter like baryons or
photons where $\delta\rho$ and $\delta p$
are quadratic in the fluctuating field. One has
\bear\label{GG6}
\delta\rho_k&=&V'\delta\chi_k+\dot\varphi\delta\dot\chi_k-V(h_{00})_k\\
\delta p_k&=&-V'\delta\chi_k+\dot\varphi\delta\dot\chi_k-
\frac{1}{3}(V-\frac{1}{2}\dot\varphi^2)(h+h_{00})_k
+\frac{1}{2}\dot\varphi^2 (h_{00})_k\nonumber\ear
For eq. (\ref{A.17}) we also need
\be\label{A.18}
\delta s_{0i}=\delta t_{0i}-\frac{1}{2}\hat T^\rho_\rho  
h_{0i}=\dot\varphi\partial_i\delta\chi
+(V-\frac{1}{2}\rho+\frac{3}{2}p)h_{0i}\ee
One observes that the rotation vector $h_{0i}$ couples to
the rotation scalars $h_{00}, h, \delta\chi$ only through the derivatives
$\partial_ih, \partial_ih_{00},\partial_i\delta\chi$ in eqs.
(\ref{A.17}), (\ref{A.18}).

We are particularly interested in the evolution of the modes with length
scale outside the horizon, $k^2/a^2\ll H^2$. For these modes
the coupling between $h_{0i}$ and $h, h_{00}, \delta\chi$ becomes
suppressed and vanishes for $k^2\to 0$. A possible solution
appears to be $h_{0i}\sim H^{-1}\partial_ih$ or
$(h_{0i})_k\sim ik_ih/H$, since
$(h_{0i})_k=0$ is indeed a solution of eq. (\ref{A.17})
for $k^2=0$. For this type
of solution an important technical simplification happens for
the long-distance modes: as compared to $h$ the quantity
\be\label{A.19}
H^{-1}\partial_ih_0^{\ i}\sim H^{-2}\frac{k^2}{a^2}h\ee
is small and may be neglected. The gauge condition (\ref{A.13})
simplifies
\be\label{A.20}
\dot h+2\dot h_{00}+2H(h+4h_{00})=0\ee
and we do not need to consider $h_{0i}$ further. The gauge condition
(\ref{A.20}) can then be used instead of the field equation
(\ref{A.15}) in order to determine the time evolution of $h_{00}$. In
particular, for $h\sim t^\lambda, h_{00}\sim t^\lambda$,
and $H=\eta t^{-1}$ one finds from (\ref{A.20})
\be\label{A.21}
\lambda h+2\lambda h_{00}+2\eta(h+4h_{00})=0\ee
This yields the ratio
\be\label{A.22}
y=h_{00}/h=-(\lambda+2\eta)/(2\lambda+8\eta)\ee
We have investigated the self-consistency of a possible
solution with $\partial_ih^i_0\approx0$  only for the case
$\lambda=0$ (see sect. 8). The question if a
solution with the property (\ref{A.19}) also exists
for $\lambda>0$ still
needs a more thorough discussion, in particular in
presence of fluctuations in the radiation.

\section*{Appendix B\protect\\
Gravitational contribution to the energy momentum tensor}  
\renewcommand{\theequation}{B.\arabic{equation}}
\setcounter{equation}{0}

In this appendix we compute the gravitational contribution to
the energy momentum tensor, eq. (\ref{2.9}). We start be evaluating
$R_{\mu\nu}$ in second order in $h_{\mu\nu}$. Using
\be\label{B1}
g^{\mu\nu}=\bar g^{\mu\nu}-h^{\mu\nu}+h^\mu_{\ \rho}h^{\rho\nu}\ee
\be\label{B2}
\Gamma_{\mu\nu}^{\ \ \lambda}=\bar\Gamma_{\mu\nu}^{\ \ \lambda}
+\frac{1}{2}(\bar g^{\lambda\rho}-h^{\lambda\rho})h_{\nu\rho
;\mu}+h_{\mu\rho;\nu}-h_{\mu\nu;\rho})\ee
one finds
\bear\label{B3}
R_{\mu\nu}&=&\bar R_{\mu\nu}+\frac{1}{2}[(\bar  
g^{\lambda\rho}-h^{\lambda\rho})(h_{\nu\rho;\mu}+h_{\mu\rho;\nu}-h_{\mu
\nu;\rho})]_{;\lambda}\nonumber\\
&&-\frac{1}{2}[(\bar  
g^{\lambda\rho}-h^{\lambda\rho})(h_{\nu\rho;\lambda}+h_{\lambda\rho;\nu}-
h_{\lambda\nu;\rho})]_{;\mu}\nonumber\\
&&+\frac{1}{4}\{h_;^\rho(h_{\mu\rho;\nu}+h_{\nu\rho;\mu}-
h_{\mu\nu;\rho})\nonumber\\
&&-(h_{\mu\ ;\lambda}^{\ \rho}+h_{\lambda\ ;\mu}^{\ \rho}-
h_{\mu\lambda;}^{\ \ \ \rho})(h_{\rho\ ;\nu}^{\ \lambda}+h_{\nu\ ;\rho}^{
\ \lambda}-h_{\rho\nu;}^{\ \ \ \lambda})\}\nonumber\\
&=&\bar R_{\mu\nu}+\frac{1}{2}(h_{\nu\rho;\mu}^{\ \ \ \ \rho}+
h_{\mu\rho;\nu}^{\ \ \ \ \rho}-h_{\mu\nu;\rho}^{\ \ \ \ \rho}-h_{;\nu\mu})
\nonumber\\
&&-\frac{1}{2} h^{\lambda\rho}(h_{\nu\rho;\mu\lambda}+h_{\mu\rho;
\nu\lambda}-h_{\mu\nu;\rho\lambda}-h_{\rho\lambda;\nu\mu})\nonumber\\
&&-\frac{1}{2}(h^{\rho\lambda}_{ \ \ ;\lambda}-\frac{1}{2}h_;^{\ \rho})
(h_{\nu\rho;\mu}+h_{\mu\rho;\nu}-h_{\mu\nu;\rho})\nonumber\\
&&+\frac{1}{4} h^{\rho\lambda}_{\ \ ;\mu}h_{\lambda\rho;\nu}-\frac{1}{2}
h_{\mu\ ;\rho}^{\ \lambda} h_{\nu\ ;\lambda}^{\ \rho}+
\frac{1}{2}h_{\mu\lambda;}^{\ \ \ \rho}h_{\nu\ ;\rho}^{\ \lambda}\ear
In lowest order only the terms quadratic in $h_{\mu\nu}$ contribute
to $<\delta G_{\mu\nu}>$:
\bear\label{B4}
<\delta G_{\mu\nu}>&=&<\delta R_{\mu\nu}-\frac{1}{2}\delta(R g_{\mu\nu})>
\nonumber\\
&=&<\delta^{(2)}R_{\mu\nu}-\frac{1}{2}\delta^{(2)}R_{\sigma\tau}
\bar g^{\sigma\tau}\bar g_{\mu\nu}
+\frac{1}{2}\delta ^{(1)}R_{\sigma\tau}(h^{\sigma\tau}\bar g_{\mu\nu}
-\bar g^{\sigma\tau}h_{\mu\nu})\nonumber \\
&&-\frac{1}{2}\bar R_{\sigma\tau}(h^\sigma_{\ \rho}h^{\rho\tau}\bar  
g_{\mu\nu}-h^{\sigma\tau}h_{\mu\nu})>\nonumber\\
&=&<-\frac{1}{2}h^{\lambda\rho}(h_{\nu\rho;\mu\lambda}+h_{\mu\rho;
\nu\lambda}-h_{\mu\nu;\rho\lambda}-h_{\rho\lambda;\nu\mu})
-\frac{1}{2} h_{\mu\nu}(h^{\lambda\rho}_{\ \ ;\lambda\rho}-
h_{;\rho}^{\ \rho})\nonumber\\
&&-\frac{1}{2}(h^{\rho\lambda}_{\ \ ;\lambda}-\frac{1}{2}h_{;}^{\ \rho})
(h_{\nu
\rho ;\mu}+h_{\mu\rho;\nu}-h_{\mu\nu;\rho})\nonumber\\
&&+\frac{1}{4} h^{\rho\lambda}_{\ \ ;\mu}h_{\lambda\rho;\nu}-\frac{1}{2}
h_{\mu\ ;\rho}^{\ \lambda} h_{\nu\ ;\lambda}^{\ \rho}+
\frac{1}{2}h_{\mu\lambda;}^{\ \ \ \rho}h_{\nu\ ;\rho}^{\  
\lambda}+\frac{1}{2}\bar R_{\rho\lambda}h^{\rho\lambda}h_{\mu\nu}>\nonumber\\
&&+\bar g_{\mu\nu}<\frac{1}{2}h^{\lambda\rho}(h_{\sigma\rho;\ \lambda}
^{\ \ \ \sigma}+h_{\sigma\rho;\lambda}^{\ \ \ \ \sigma}-
h_{;\rho\lambda}-h_{\rho\lambda;\sigma}^{\ \ \ \ \sigma})
\nonumber\\
&&+\frac{1}{2}(h^{\rho\lambda}_{\ \ ;\lambda}-\frac{1}{2}h_;^{\ \rho})
(h_{\rho\ ;\sigma}^{\ \sigma}-\frac{1}{2}h_{;\rho})+\frac{1}{4}h_{\sigma\  
;\rho}^{\ \lambda}h^{\sigma\rho}_{\ \ ;\lambda}
-\frac{3}{8}h_{\sigma\lambda;}^{\ \ \ \rho}h^{\sigma\lambda}_{\ \  ;\rho}>
\ear
This expression can be somewhat simplified by use of the harmonic
gauge condition
\bear\label{B5}
\delta G_{\mu\nu}&=&\frac{1}{2}<h^{\lambda\rho}(h_{\mu\nu;\rho\lambda}
+h_{\rho\lambda;\mu\nu}-h_{\nu\rho;\mu\lambda}-h_{\mu\rho;\nu\lambda})
\\
&&+\frac{1}{2}h_{\mu\nu}h_{;\rho}^{\ \ \rho}+\frac{1}{2} h^{\rho\lambda}_{\ \  
;\mu}h_{\lambda\rho;\nu}-h^{\ \lambda}_{\mu\ ;\rho}h_{\nu \ ;\lambda}^{\ \rho}
+h_{\mu\lambda;}^{\ \ \ \rho}h^{\ \lambda}_{\nu\ ;\rho}+\bar  
R_{\rho\lambda}h^{\rho\lambda}h_{\mu\nu}>\nonumber\\
&&-\frac{1}{2}\bar g_{\mu\nu}<h^{\lambda\rho}h_{\rho\lambda;\sigma}^{\ \ \ \
\sigma}-\frac{1}{2}h_{\sigma\ ;\rho}^{\ \lambda}h^{\sigma\rho}_{\ \ ;\lambda}
+\frac{3}{4}h_{\sigma\lambda;}^{\ \ \ \rho}h^{\sigma\lambda}_{\ \ ;\rho}
-h^{\lambda\rho}(\bar R_{\lambda\tau}h^\tau_{\ \rho}-\bar R
_{\lambda\sigma\rho\tau}h^{\sigma\tau})>\nonumber\ear
Eq. (\ref{2.9}) yields the gravitational energy density
$\rho_g$ and pressure $p_g$.

We are interested here in the contribution from fluctuations with wavelength
larger than the horizon\footnote{ See \cite{CWBR} for a discussion
of the wavelengths inside the horizon.}.
We therefore can neglect the space derivatives
and approximate $h_{0i}=0$. Inserting the Robertson-Walker background
one finds
\bear\label{B6}
\delta G_{\mu\nu}&=&\frac{1}{2}\bar g_{\mu\nu}<\frac{1}{3}h(\ddot h+\ddot h_{00})
+\frac{1}{3}h_{00}(\ddot h+4\ddot h_{00})+
\frac{1}{4}\dot h^2_{00}+\frac{3}{4}\dot h^j_i\dot h_j^i\nonumber\\
&&+H\left\{\frac{5}{3}h\dot h+\frac{7}{3}h\dot h_{00}+\frac{14}{3}
h_{00}\dot h+\frac{31}{3}h_{00}\dot h_{00}+\frac{1}{2} h^i_j\dot h_j^i
\right\}\nonumber\\
&&+H^2\left\{-\frac{1}{3}h^2+\frac{19}{3}hh_{00}+\frac{103}
{6}h^2_{00}+\frac{19}{4}h_i^jh^i_j\right\}
\nonumber\\
&&+\dot H\left\{2hh_{00}+5h^2_{00}+h_i^jh_j^i\right\}>+\delta F_{\mu\nu}
\ear
with $\delta F_{\mu\nu}$ a similar
contribution from the first bracket in eq. (\ref{B5}). For the constant
solution, $h_{00}=-\frac{1}{4}h,\  \dot h
=0$, a radiation background $\dot H=-2H^2$, and using
$h_{ij}=\frac{1}{3}h_i^{\ i}g_{ij}=\frac{1}{3}(h+h_{00})g_{ij}$
one obtains\footnote{We have not yet computed the proportionality
factor $c$ which could vanish.}
\be\label{B7}
(\rho_g)_k=\frac{c}{8}M^2H^2<h^2>_k=\frac{c\alpha^2}{2t^2}\alpha_k\ee

\section*{Appendix C\protect\\
Naming scheme for energy densities}
\renewcommand{\theequation}{A.\arabic{equation}}
\setcounter{equation}{0}

In this appendix we summarize our naming scheme for the different
contributions to the energy density $\rho_j$, and similar for
$p_j$ and $\Omega_j$:\\

\bigskip
$\rho_h$\ :\ homogenous quintessence; smoothly distributed over
distances $z\leq 10$.

\bigskip
$\rho_c$\ :\ cosmon; contribution of scalar fluctuations with wavelength
$z\leq 10$.

\bigskip
$\rho_\varphi$\ :\ scalar; $\rho_\varphi=\rho_h+\rho_c$

\bigskip
$\rho_g$\ :\ gravitational

\bigskip
$\rho_q$\ :\ quintessence; $\rho_q=\rho_\varphi+\rho_g$

\bigskip
$\lambda$\ :\ cosmological constant $(\rho_\Lambda\equiv\lambda$)

\bigskip
$\rho_{de}$\ :\ dark energy; $\rho_{de}=\rho_h+\lambda$

\bigskip
$\rho_{cdm}$\ :\ cold dark matter

\bigskip
$\rho_{dm}$\ :\ dark matter; $\rho_{dm}=\rho_c+\rho_g+\rho_{cdm}(+\rho_\nu+...)$

\bigskip
$\rho_d$\ :\ dark; $\rho_d=\rho_{de}+\rho_{dm}$.

\bigskip
$\rho_b$\ :\ baryons;\qquad $\rho_\nu$\ :\ neutrinos;
\qquad $\rho_r$\ :\
radiation

\bigskip
$\rho_m$\ :\ nonrelativistic matter;\quad
$\rho_m=\rho_{cdm}+\rho_b(+\rho_\nu+...)$

\bigskip
$\rho_p$\ :\ particles;\quad $\rho_p=\rho_{cdm}+\rho_r+\rho_\nu+\rho_b+...$

\bigskip
$\rho_{inhom}$\ :\ inhomogenous; takes part in structure formation;
$\rho_{inhom}=\rho_c+\rho_g+\rho_m$\\
\phantom{$\rho_{inhom}$\ :\ }\quad\  (we often use $\rho_{inhom}\to\rho$)

\bigskip
$\rho_{tot}$\ :\ total; $\rho_{tot}=\rho_{de}+\rho_{inhom}+\rho_r+...$

\bigskip\noindent
An interesting candidate for recent and present cosmology would be
a universe consisting mainly of homogenous quintessence and
cosmon dark matter, $\rho_{tot}=\rho_h+\rho_c+\rho_g=\rho_q=\rho_h+\rho_{inhom}$.

\end{document}